\documentstyle[aaspp4,epsf,lscape]{article}

\def\hr{\hbox{${}^{\hbox{\small h}}$}}
\def\mn{\hbox{${}^{\hbox{\small m}}$}}
\def\sc{\hbox{${}^{\hbox{\small s}}$\llap{.}}}
\def\deg{\hbox{${}^\circ$}}
\def\min{\hbox{${}^{\prime}$}}
\def\sec{\hbox{${}^{\prime\prime}$}}
\begin{document}

\title{The {\sl Hubble Space Telescope} Extragalactic Distance Scale Key 
Project. X. The Cepheid Distance to NGC~7331\footnote{Based on  
observations   with  the NASA/ESA   {\it  Hubble Space
Telescope}, obtained at the Space  Telescope Science  Institute, which
is operated by AURA, Inc. under  NASA Contract No.  NAS 5-26555.}}

\author{Shaun~M.~G.~Hughes,\footnote{Royal Greenwich Observatory, 
Madingley Road, Cambridge CB3 0EZ, UK; hughes@ast.cam.ac.uk} 
Mingsheng~Han,\footnote{University~of~Wisconsin, Madison, Wisconsin~53706, USA;
mhan@malachi.gsfc.nasa.gov} 
John~Hoessel,\footnote{University~of~Wisconsin, Madison, Wisconsin~53706, USA;
hoessel@uwfpc.astro.wisc.edu} 
Wendy~L.~Freedman,\footnote{The~Observatories,  
Carnegie~Institution~of~Washington, Pasadena, CA~91101; wendy@ociw.edu} 
Robert~C.~Kennicutt,~Jr.,\footnote{Steward~Observatory,  University~of~Arizona,
Tucson, AZ~85721; robk@as.arizona.edu} 
Jeremy.~R.~Mould,\footnote{Mount~Stromlo~\&~Siding~Spring~Observatories, 
Australian~National~University, Canberra, Australia; jrm@merlin.anu.edu.au} 
Abi~Saha,\footnote{Space~Telescope~Science~Institute,  3700~San~Martin~Drive,
Baltimore, MD~21218; saha@stsci.edu} 
Peter~B.~Stetson,\footnote{Dominion~Astrophysical~Observatory, 
5071~W.~Saanich~Rd., Victoria~BC~V8X~4M6; peter.stetson@hia.nrc.ca} 
Barry~F.~Madore,\footnote{NASA/IPAC~Extragalactic~Database, 
Infrared~Processing~and~Analysis~Center, California~Institute~of~Technology,  
Pasadena, CA~91125; barry@ipac.caltech.edu} 
Nancy~A.~Silbermann,\footnote{Infrared~Processing~and~Analysis~Center, 
Jet~Propulsion~Laboratory, California~Institute~of~Technology, Pasadena, 
CA~91125; nancys@ipac.caltech.edu} 
Paul~Harding,\footnote{Steward~Observatory, University~of~Arizona, Tucson, 
AZ~85721; harding@as.arizona.edu} 
Laura~Ferrarese,\footnote{Palomar~Observatory, 
California~Institute~of~Technology, Pasadena, CA~91125; lff@astro.caltech.edu} 
Holland~Ford,\footnote{Dept~of~Physics~\&~Astronomy, Bloomberg~501,
Johns~Hopkins~Univ.,  3400~N.~Charles~St.,  Baltimore, MD~~21218;
ford@stsci.edu} 
Brad~K.~Gibson,\footnote{Mount~Stromlo~\&~Siding~Spring~Observatories, 
Australian~National~University, Canberra, Australia; gibson@mso.anu.edu.au} 
John~A.~Graham,\footnote{Department~of~Terrestrial~Magnetism,  
Carnegie~Institution~of~Washington, 5241~Broad~Branch~Rd.~N.W., 
Washington~D.C.~20015; graham@jag.ciw.edu} 
Robert~Hill,\footnote{Laboratory~for~Astronomy~\&~Solar~Physics, 
NASA~Goddard~Space~Flight~Center, Greenbelt, MD~20771; hill@esa.nascom.nasa.gov} 
John~Huchra,\footnote{Harvard-Smithsonian~Center~for~Astrophysics, 
60~Garden~Street, Cambridge, MA~02138; huchra@cfa.harvard.edu} 
Garth~D.~Illingworth,\footnote{Lick~Observatory, University~of~California, 
Santa~Cruz, CA~95064; gdi@ucolick.org} 
Randy~Phelps,\footnote{The~Observatories, Carnegie~Institution~of~Washington, 
Pasadena, CA~91101; phelps@ociw.edu} 
Shoko~Sakai \footnote{Infrared~Processing~and~Analysis~Center, 
Jet~Propulsion~Laboratory, California~Institute~of~Technology, Pasadena, 
CA~91125; shoko@ipac.caltech.edu}}

\vspace{0.6cm}
\centerline{Accepted for publication in the Astrophysical Journal, vol 501, 
1998 July 1}

\newpage

\begin{abstract}
The distance to NGC~7331 has been derived from Cepheid variables
observed with HST/WFPC2, as part of the Extragalactic Distance Scale Key 
Project.  Multi-epoch exposures in F555W ($\sim V$) and F814W ($\sim I$), 
with photometry derived independently from DoPHOT and DAOPHOT/ALLFRAME
programs, were used to detect a total of 13 reliable Cepheids, with 
periods between 11 and 42 days.  The relative distance moduli between 
NGC~7331 and the LMC, derived from the $V$ and $I$ magnitudes, imply 
an extinction to NGC~7331 of A$_V = 0.47\pm0.15$ mag, and an 
extinction-corrected distance modulus to NGC~7331 of 
$30.89\pm0.14$(random) mag, equivalent to a distance of 
$15.1 ^{+1.0}_{-0.9}$ Mpc.  There are additional systematic uncertainties
in the distance modulus of $\pm0.12$ mag due to the calibration of the 
Cepheid Period-Luminosity relation, and a systematic offset of 
$+0.05\pm0.04$ mag if we applied the metallicity correction inferred 
from the M101 results of Kennicutt et al 1998.  
\end{abstract}


\section{Introduction}

The {\sl Hubble Space Telescope (HST)} Extragalactic Distance Scale Key 
Project aims to obtain Cepheid distances to 18 galaxies within 20 Mpc, 
to use these to calibrate secondary distance
estimators, and thereby measure H$_0$ to an external accuracy of 10\%
(\markcite{kenn95}Kennicutt, Freedman \& Mould 1995).
NGC~7331 (Figure~\ref{INT_fig}) was chosen primarily as a calibrator for
the luminosity-line width relation (also known as the Tully-Fisher relation).
It lies in the constellation of Pegasus, at position 
$\alpha =$ 22\hr37\mn05\sc2, $\delta =$ +34\deg25\min10\sec\ (J2000), 
and has a Galactocentric velocity of +1035 km/s 
(\markcite{deva91}de Vaucouleurs et al.\ 1991; hereinafter RC3).  

NGC~7331 has an early type spiral classification of Sb(rs)I-II in the Revised 
Shapley-Ames scheme by \markcite{sand81}Sandage \& Tammann (1981), a Hubble 
type T = 3 in \markcite{deva91}RC3, and an inclination of $\sim$ 75\arcdeg\ 
(estimates vary from 71\arcdeg\ by \markcite{huch89}Huchtmeier \& Richter 1989 
to $75 \pm 5$\arcdeg\ by \markcite{marc94}Marcelin et al.\ 1994 and 
$74.8 \pm 2.0$\arcdeg\ by \markcite{bege87}Begeman 1987).
The \ion{H}{2} rotation curve for NGC~7331 is very regular, and peaks at
$\sim$250 km/s (\markcite{rubi65}Rubin et al.\ 1965), and the  
\ion{H}{1} (21 cm) linewidth ($W_{20}$) has been measured as $531 \pm 10$ km/s 
(\markcite{fish81}Fisher \& Tully 1981) and 536 km/s (from 
\markcite{bege87}Begeman 1987).  
The properties of being a regular spiral of moderately large inclination 
make NGC~7331 an ideal calibrator for the Tully-Fisher relation.

We present here the distance to NGC~7331, using observations obtained
with the post-repair mission Wide Field and Planetary Camera (WFPC2) on HST  
with the F555W and F814W filters (see \markcite{holt95a} Holtzman et al.\ 
1995a, and the current version of the WFPC2 Instrument Handbook from STScI, 
for details of the performance of WFPC2).  Our methodology follows that 
used in previous papers in this series: e.~g.\ 
M81 (\markcite{free94a}Freedman et al.\ 1994a), 
M100 (\markcite{free94b}Freedman et al.\ 1994b and 
\markcite{ferr96}Ferrarese et al.\ 1996), 
M101 (\markcite{kels96}Kelson et al.\ 1996), 
NGC~925 (\markcite{silb96}Silbermann et al.\ 1996), 
NGC~2090 (\markcite{phel98}Phelps et al.\ 1998),
NGC~3351 (\markcite{grah97}Graham et al.\ 1997), and 
NGC~3621 (\markcite{raws97}Rawson et al.\ 1997).
Photometry for all epochs was obtained with programs designed for crowded 
fields (\S~\ref{Photometry}).  The resultant lightcurves were then searched
to identify and measure periods and luminosities of 13 Cepheid variables 
(\S~\ref{Cepheids}), applying the 
calibrations described in \markcite{ferr96}Ferrarese 
et al.\ (1996) and \markcite{hill98}Hill et al.\ (1998).
Period-luminosity relations were fitted to these Cepheids, relative to the 
Large Magellanic Cloud (LMC), to derive extinction-corrected distance moduli 
for each photometry set (\S~\ref{Distance}).
Implications for the properties of NGC~7331, and previous calibrations of the 
Tully-Fisher (TF) relation (\S~\ref{Disc}) are briefly discussed.
Preliminary results of this work were described in 
\markcite{hugh96}Hughes, Han \& Hoessel (1996).  

\section{Photometry}
\label{Photometry}

The WFPC2 field lies 3.5 arcmin north of the galaxy's nucleus, along the major 
axis (Figures~\ref{INT_fig} and ~\ref{wfpc2_cephs}).  Variable stars were 
detected from 15 epochs of F555W cosmic ray-split exposures, and their colors 
were measured from 4 epochs of F814W exposures, over the interval of 
1994 June 18 to 1995 August 17, as listed in Table~\ref{dates_tab}.  Each 
cosmic ray-split exposure pair consisted of a 1600 sec and a 1200 sec exposure 
(a safing event part way through the series meant that three F555W and one 
F814W epochs had to be postponed, and resulted in missed 1600 sec exposures 
for F555W at epoch 7 and F814W at epoch 8).  In all cases the gain was 
7 e$^-/$DN, with a readout noise of 7 e$^-$.  All of these epochs were taken 
after the decrease in the WFPC2 operating temperature (which occurred on 
1994 April 23).  Although there appears to be a charge transfer efficiency 
effect in the WFPC2 CCDs, this may be significant only in exposures with a 
low background, and seems to be negligible for backgrounds brighter than 
70e$^-$ (\markcite{raws97}Rawson et al.\ 1997).  As our NGC~7331 frames
all have backgrounds brighter than 70e$^-$, they should not have been affected.
Two short exposures of 260 sec in F555W and F814W were also made at epoch 13, 
but have not been used as they were intended for calibration backup via 
ground-based observations in case there were any problems with the photometric 
calibrations. 

The science frames were processed through the standard STScI
pipeline (\markcite{holt95a}Holtzman et al.\ 1995a).  The photometric
distortions introduced by the WFPC2 corrective optics were rectified
by multiplying each image by a pixel area correction map.  These differ
slightly from those used by \markcite{holt95b}Holtzman et al.\ (1995b) in that 
the normalization is relative to the median pixel, rather than the largest.

Various pixels were systematically ignored by the photometry programs:
those which are off the WFPC2 pyramid surface; or vignetted by the
pyramid edge; or given a non-zero value in the bad pixel maps.    

The photometry was derived using two independent programs, in order to
identify and trap any possible mistakes in the reduction process by comparing
their final results.  The first program is based on DoPHOT 
(\markcite{sche93}Schechter, Mateo \& Saha 1993;
\markcite{saha94}Saha et al.\ 1994; \markcite{saha96}Saha et al.\ 1996), and 
the second is DAOPHOT/ALLFRAME (\markcite{stet94}Stetson 1994), used in a 
similar way as described in detail for M100 by \markcite{ferr96}Ferrarese 
et al.\ (1996) and \markcite{hill98}Hill et al.\ (1998), adopting the long 
exposure calibration (see Appendix).
  
\subsection{DoPHOT}

DoPHOT measures magnitudes by fitting a purely analytic model point-spread
function (PSF) to objects detected above a user-specified threshold.
For DoPHOT, we followed the methodology described in \markcite{saha96}Saha
et al. (1996; sections 2 and 3),\footnote{With the exception that here we have 
corrected for the photometric distortions of the corrective optics by the 
multiplication of the pixel area map.} which we will only briefly describe 
here.  Each cosmic ray-split pair of observations was first combined
using a sigma detection algorithm which takes into account the problems of
undersampling (see \markcite{saha96}Saha et al.\ 1996 for a full description),
to produce single frames per epoch mostly free of cosmic rays.
The cosmic ray-cleaned images were run through DoPHOT, to obtain the positions 
of the brightest stars, which were then matched (using DAOMATCH and DAOMASTER, 
\markcite{stet92}Stetson 1992) to obtain the coordinate transformations 
between epochs.  The X and Y offsets were all less than $\sim 0.2$ arcsec.  
All the frames for each chip were then shifted by their offsets and 
median-combined to produce clean and 
reasonably deep master images, which were then passed through DoPHOT, 
to produce a master list of objects, retaining only those objects DoPHOT 
classified as stars.  The coordinates of the master list were then transformed 
to the positions of each epoch, and each of these lists were used within 
DoPHOT to obtain raw DoPHOT photometry of each of the master list objects in 
each of the epoch frames.  

The raw photometry was then calibrated to equivalent 0.5 arcsec aperture 
magnitudes, using aperture corrections and  measured zero-points (appropriate
for long exposures) for both filters and each chip (see Appendix A).  The 
photometry of all epochs for 
each object, now calibrated to 0.5 arcsec aperture magnitudes,  was then 
combined using DAOMASTER, to produce a light curve for every object in the 
input master list.  

\subsection{ALLFRAME}

For ALLFRAME, the original science frames were used (i.~e.\ treating each of 
the cosmic ray-split exposures as two individual epochs), following the 
ALLFRAME method described for previous galaxies in this series 
(e.~g.\ \markcite{ferr96}Ferrarese et al.\ 1996; 
\markcite{silb96}Silbermann et al.\ 1996).  The DAOPHOT/ALLFRAME program 
(\markcite{stet94}Stetson 1994) takes a different approach in doing the 
photometry, in that an analytic model of the PSF is augmented by an 
empirically-determined difference map, which represents the intensity-weighted 
mean difference between the observed PSF of a large sample of isolated stars 
and the analytic function.  This model of the PSF is also allowed to vary as 
a quadratic function of position on the chip, to mimic the variations in the 
true PSF across each chip.  In the case of the HST images of NGC~7331, there 
were too few isolated stars to adequately characterize the PSF, so the model 
PSFs derived from a number of independent images of globular cluster fields 
were used.  ALLFRAME then fits these model PSFs (one per filter and chip) to 
all stars on all the frames. To do this, ALLFRAME must first have a master 
input list of objects, which like DoPHOT, was derived from median images 
produced from coordinate transformations defined by matches of bright isolated 
stars on each of the individual 
cosmic ray-cleaned frames.  Unlike DoPHOT, ALLFRAME only 
uses this star list as a starting point, and uses the results of simultaneous 
PSF fitting on all the frames to improve the coordinate transformations 
between frames, thereby iteratively improving the master list.  The final 
photometry files produced by ALLFRAME for each epoch were then matched 
using DAOMASTER, to produce a single file containing the positions and 
magnitudes of all the matched stars at all the epochs.

Although cosmic rays are left in each of the single epoch frames, these
are generally not a problem, as the position of each star is determined
solely from the master list and the coordinate transformation of the entire 
frame.  Thus cosmic rays falling near a star will not affect its centroid,
and are rejected as being deviant pixels during the PSF fitting.  The only
time a cosmic ray will affect the photometry is when it impacts
squarely on the peak of the stellar profile, making it appear that the star 
is unusually bright in that one image.  Such defects must be identified and 
removed at a later stage of the analysis, by comparison with the rest of
the lightcurve.

The aperture corrections were determined for each epoch, to allow for focus 
variations (since unlike DoPHOT, the model PSF was fixed in ALLFRAME).  As
there were few isolated stars to do this, the procedure was made more robust 
by determining a mean set of growth curves for the observed PSFs of isolated 
stars in each chip/filter from all the Key Project galaxies to date, which 
were used to derive the 0.5 arcsec aperture magnitudes for the few isolated 
stars, then measuring the mean epoch aperture correction as the difference 
between this aperture magnitude and the PSF (ALLFRAME) magnitude. 
The rms dispersion about the mean for these aperture corrections were 
less than $\sim$0.01 magnitude, except for F814W chip 3, which was 0.02 
magnitude.  This scatter is insignificant compared to the uncertainties of the 
individual PSF fits of 0.1 to 0.3 magnitude which are typical of the 
Cepheids.   The mean aperture corrections for each epoch/chip/filter, as well 
as zero-points appropriate for long exposures, are listed in 
Appendix A.  These were added to the PSF magnitudes of all objects, to obtain 
0.5 arcsec magnitudes (F555W$_{0.5}$ and F814W$_{0.5}$). 

\subsection{Comparison Between DoPHOT and ALLFRAME Photometry}

To allow easier external comparisons with our calibrated magnitudes, 
we list in Table~\ref{Ref_mags_tab} the positions and calibrated mean 
ALLFRAME and DoPHOT magnitudes ($V$ and $I$) of stars which are isolated 
and unaffected by cosmic ray events, and which DoPHOT classified as stars.  
Due to crowding, there are not as many of these isolated reference stars as 
we would have liked.

The calibrated magnitudes (i.~e.\ $V$ and $I$) differences of these
(reference) stars are given in Table~\ref{Mags_comp_tab}, and plotted in 
Figure~\ref{doC_alf_comp_fig}.  In the wide field chips (chips 2-4),
where the Cepheids are, these reference star magnitudes agree to within 
0.05 mag.  
Some of the scatter between these DoPHOT and ALLFRAME magnitudes arises
from the different methods used to fit the model PSF, in that DoPHOT weights 
each pixel independently by its own noise estimate, whereas ALLFRAME 
weights relative to its PSF profile.  Although the comparison 
of the Cepheid photometry itself yields better agreement between the two 
programs, we are continuing to try and understand the source of these 
remaining differences. In particular, we are undertaking a series of 
artificial star experiments designed to investigate the effects of crowding,
differences in PSF's, the differences in the way the two programs
calculate sky values, etc.

An even more relevant comparison, in terms of the distance to NGC~7331, 
is the difference in the calibrated Cepheid magnitudes (see \S~\ref{Cepheids} 
and \S~\ref{Distance}), which is reflected in the difference between the 
Period-Luminosity fits in Table~\ref{PL_fit_tab}.  This indicates that the 
DoPHOT and ALLFRAME calibrated magnitudes agree to $\sim$0.05 mag in $V$, 
a difference of less than one sigma with respect to 
the internal errors, and well within the external errors.
In $I$, the agreement is even better.
The difference between the magnitudes for the Cepheids are also plotted in
Figure~\ref{doC_alf_comp_fig} (open symbols).

\section{The Cepheids}
\label{Cepheids}

To detect the variables, the objects in the DoPHOT photometry sets were first 
grouped into magnitude bins of width 0.5 mag.  A variable was then defined
as any object with a robust rms-like dispersion (the F-$\sigma$\
parameter, based on the inner range of variability, as described in 
\markcite{hugh89}Hughes 1989, and derived from \markcite{hoag83}Hoaglin, 
Mosteller \& Tukey 1983) greater than twice the mean rms of its magnitude 
bin. Each object was also required to have been matched in at least 10 epochs.
These variables were then searched for likely periods between 10 and 80 days, 
using a program based on the Phase Dispersion Minimization (PDM) algorithm of 
\markcite{stel78}Stellingwerf (1978), which is based on the 
\markcite{lafl65}Lafler \& Kinman (1965) statistic.  The main advantage of 
this method is that it is efficient at detecting periodicity in small epoch 
samples, and is insensitive to light curve shape.  Simultaneously, the image 
of each variable was viewed, to ensure it was not a spurious variable caused 
by a close association with another object, as well as its light curve, 
to ensure no spurious objects were selected due to noise or cosmic ray events.

For the ALLFRAME photometry sets, variable candidates were identified using 
two indicators.  The first is similar to the $\chi ^2$\ parameter of 
\markcite{saha90}Saha \& Hoessel (1990), and the second is the color index 
parameter of \markcite{welc93}Welch \& Stetson (1993).  Cepheid candidates 
were identified by visually examining the phased light curves of the variable 
candidates.  Any epochs contaminated by cosmic rays were removed interactively.

The final list of 13 good Cepheids was selected on  the basis that each had 
to appear Cepheid-like in both candidate lists: a Cepheid-like lightcurve 
(sharp rise, longer decay for those with periods $>$ 20 days, more sinusoid 
for periods $<$ 20 days); similar periods; similar amplitudes; and colors
consistent with being in the instability strip.      No Cepheids were found 
on chip 1, the planetary camera (PC) chip.  This is mainly due to the
smaller field of view, for which we would expect just 1.1 Cepheids
(based on the mean number of Cepheids found in the WF chips, assuming the
stellar population is uniform across all chips).  We would also 
expect to find fewer Cepheids on the PC chip as it has smaller pixels, hence it
has larger photometric errors at the magnitudes of the Cepheids, compared to 
the other wide field (WF) chips (errors of 0.25 to 0.6 mag, at magnitudes of 
25 to 27 in $V$, compared to errors of 0.1 to 0.3 mag on the WF chips), 
resulting in much lower yields of variable candidates due to the 2-$\sigma$\ 
selection.  For Cepheids to have made this cut, they would need to have 
amplitudes greater than $\sim1.0$ mag, and only half the Cepheids in 
Table~\ref{good_ceph_tab} have such large amplitudes.  The F555W$_{0.5}$ and 
F814W$_{0.5}$ magnitudes at each epoch are listed in Table~\ref{good_cephotD} 
(DoPHOT) and Table~\ref{good_cephotH} (ALLFRAME, where the  cosmic ray-split 
epoch magnitudes are intensity averaged).

As in other papers in this series, we calculate two estimates of the
mean magnitudes of our Cepheids: mean F555W$_{0.5}$\ magnitudes derived from 
unweighted  intensity averaged means:
\begin{equation}
\langle {\rm F555W}_{0.5} \rangle = -2.5{\rm log}_{10}(\frac{1}{N}\Sigma 10^{-0.4m_i})
\end{equation}
and phase-weighted magnitudes 
\begin{equation}
({\rm F555W}_{0.5})_\phi = -2.5{\rm log}_{10}(\frac{1}{N}\Sigma 0.5(\phi_{i+1}-\phi_{i-1})
10^{-0.4m_i}) 
\end{equation}

Since the amplitudes of Cepheids decrease with wavelength, fewer $I$-band than 
$V$-band observations are required to determine mean magnitudes to the same
accuracy.  The mean 
$I$-band magnitudes were estimated based on 4 epochs following the method 
developed by \markcite{free88}Freedman (1988) and applied by 
\markcite{ferr96}Ferrarese et al.\ (1996).  This uses the fact that the 
amplitude in $V$\ is almost twice the amplitude in $I$, hence a well-sampled 
mean $I$\ magnitude $\langle I \rangle _{12}$ (over $\sim 12$ epochs) will be 
estimated as 
\begin{equation}
\langle I \rangle _{12} = \langle I \rangle _4 + 0.51(\langle V \rangle _{12}
 - \langle V \rangle _4)
\end{equation}
where $\langle I \rangle _4$\ and $\langle V \rangle _4$\ are the intensity
averaged $I$ and $V$ magnitudes at those epochs (up to four) where both F555W 
and F814W images were obtained. 

We used the optimized sampling algorithm developed by \markcite{free94a}Freedman 
et al.\ (1994a) to choose epochs which obtain good phase coverage over all bright 
Cepheid periods (i.e. periods between 10 and 60 days), hence there is very little 
difference between these mean and phase-weighted magnitudes.  Maintaining
consistency with previous papers in this series, we will use only the 
phase-weighted magnitudes in deriving distances, and these are listed in 
Table~\ref{good_ceph_tab}, where the F555W$_{0.5}$ and F814W$_{0.5}$ magnitudes
have been converted to Johnson $V$ and Kron-Cousins $I$ using the 
\markcite{holt95b}Holtzman et al.\ (1995b)
transformations in Appendix A.

In Table~\ref{good_ceph_tab} we list the parameters for the final sample of 
13 Cepheids: ID; WFPC2 chip number; position (both the RA and Dec at equinox 
J2000, and their X and Y pixel position on each chip); ALLFRAME period;
ALLFRAME phase-weighted $V$ magnitude; ALLFRAME $V-I$ color; 
DoPHOT period; DoPHOT phase-weighted $V$ magnitude; and DoPHOT $V-I$ color.
The Cepheids are identified in the WFPC2 field in Figure~\ref{wfpc2_cephs}, 
as well as in more detailed finding charts in Figure~\ref{FC_fig}, and their 
light curves, phased to the period given in Table~\ref{good_ceph_tab}, are 
shown in Figure~\ref{lc_fig}. Although V4 appears crowded, and V8 is 
near a bright star (see Figure~\ref{FC_fig}), they are still retained as 
their photometric uncertainties are no larger than other stars of similar
magnitude, and the near neighbors subtracted cleanly after PSF fitting.
The mean difference between the periods of the 
two DoPHOT and ALLFRAME Cepheid data sets $\Delta$Period $= -0.03$ days, with 
an rms dispersion of 1.27 days.
 
The Cepheids in Table~\ref{good_ceph_tab} are marked in the deep 
$I$ vs $V-I$ color-magnitude diagram (CMD)
in Figure~\ref{CMD_fig}, derived
from the mean magnitudes from all epochs (for clarity, only every 10th star 
in the CMD is plotted). All of the  Cepheids lie in the instability 
strip.  We also plot the locus of the blue plume, red supergiant and red giant 
branch for Shapley Constellation III, shifted in magnitude and color to match 
the relative distance and extinction of NGC~7331.  The position of the red 
giant branch locus implies our limiting magnitude is too bright to apply the 
tip of the red giant branch distance method (e.~g.\ \markcite{lee93}Lee et 
al.\ 1993) as a check on our Cepheid distance.

The luminosity function of the blue plume stars (i.~e.\ with $V-I < 0.3$) is 
plotted in Figure~\ref{LF_fig}.  Before incompleteness (at $V > 26$), the 
slope of the luminosity function is 0.61, similar to that of M81 (0.57; 
\markcite{hugh94}Hughes et al.\ 1994), and consistent with other late-type 
galaxies (\markcite{free85}Freedman 1985).  The luminosity function for stars 
with Cepheid-like colors ($0.5< V - I <1.5$) is also plotted in 
Figure~\ref{LF_fig}, and shows that incompleteness for Cepheids sets in at 
$V > 27.8$ mag, consistent with our faintest Cepheid in 
Table~\ref{good_ceph_tab}, which has $\langle V \rangle = 27.4$ (V8).  
The degree to which crowding and limiting magnitude affects Cepheid 
completeness is being investigated via artificial star tests by 
Ferrarese et al.\ (in preparation).

\section{The Distance to NGC~7331}
\label{Distance}

To obtain a Cepheid distance, we follow the methodology of previous papers in 
this series, and fit $V$ and $I$ PL relations relative to the LMC Cepheids, 
adopting the distance and reddening to the LMC of $18.5\pm0.1$ mag and 
$E(V-I) = 0.13$ (i.~e.\ $E(B-V) = 0.10, A_V = 3.3E(B-V)$, and use the 
\markcite{card89}Cardelli et al.\ 1989 extinction law 
$A_I/A_V = 0.7712 - 0.5897/R_V = 0.592$ for $R_V = 3.3$) from 
\markcite{mado91}Madore \& Freedman (1991; hereafter MF91),
and fitting the MF91 PL relations, derived from a set of observations of 32 
LMC Cepheids:

\begin{equation}
M_V = -2.76(\pm0.11)[{\rm log}(P) - 1.4] - 5.26(\pm0.05);  ~~~~ rms=0.27
\end{equation}

\begin{equation}
M_I = -3.06(\pm0.07)[{\rm log}(P) - 1.4] - 6.09(\pm0.03);  ~~~~ rms=0.18
\end{equation}

\noindent to obtain $V$ and $I$ distance moduli ($\mu_V, \mu_I$).
\footnote{\markcite{tanv96}Tanvir (1996) has re-derived the LMC
PL relations, from a compilation of data from several sources, of 
53 LMC Cepheids, and finds they are best fit by PL relations of the form:

$M_V = -2.774(\pm0.083)[{\rm log}(P) - 1.4] - 5.262(\pm0.040); ~~~~  rms=0.233$

$M_I = -3.039(\pm0.059)[{\rm log}(P) - 1.4] - 6.049(\pm0.028); ~~~~  rms=0.164$

There is very little difference in the $V$ PL relation, but a significant
difference in the $I$ PL relation zero-point.  \markcite{tanv96}Tanvir has
attributed this difference to the fact that for the Cepheids in 
\markcite{mart75}Martin, Warren \& Feast (1979), MF91 used $I$ magnitudes 
derived from $\langle V \rangle - \overline{V-I}$ (where $\overline{V-I}$ is 
the mean of the $V-I$ colors at all epochs), rather than intensity means of 
the $I$ data. For Cepheids of normal amplitudes, this creates systematic 
errors of 0.02-0.09 mag.  If confirmed, this means that use of the 
MF91 relations would produce a systematic overestimate of the relative 
$I$ distance modulus of a galaxy by $\sim 0.04$ mag, hence of the reddening, 
which produces an overestimate of the extinction-corrected distance modulus of 
0.1 mag, and a corresponding underestimate of H$_0$ by 5\%.  An ongoing 
program to measure mean $I$-band magnitudes for a sample of LMC Cepheids 
being carried out by us at Las Campanas Observatory and Mount Stromlo will 
shed further light on this issue.  In the meantime, to be consistent with 
previous papers in this series, we continue to use the MF91 PL relations.}

As can be seen in the image of Figure~\ref{wfpc2_cephs} (and to a more 
dramatic extent in Figure~\ref{INT_fig}), there are noticeable amounts of 
dust in NGC~7331, so it is important to measure and account for this as 
accurately as possible.  As in MF91, we measure the mean reddening $E(V-I)$ 
for the whole sample of Cepheids from the difference $\mu_V - \mu_I$, which
is then used to derive the extinction-corrected distance.  

We have two methods of measuring apparent distance moduli in $V$ and $I$.  
The first adopts the LMC PL relations of MF91, using least squares fits with
slopes fixed at the MF91 values.  The second method is that used by 
\markcite{ferr96}Ferrarese et al.\ (1996) for M100.  This assumes nothing
about the slopes of PL relations, but simply adds a common magnitude offset
to the set of LMC Cepheids with periods greater than 10 days used by MF91, 
such that when their magnitudes are combined with the NGC~7331 Cepheids,
the scatter of the combined set is minimized.  The final offset is then
the difference in apparent distance modulus between the LMC and NGC~7331.
Both methods produced virtually 
identical results, and so we only present the results of the first method.  
The PL relations are shown in Figures~\ref{PL_figH} and \ref{PL_figD},
where the fits using the MF91 slopes are the solid
lines, with the dotted lines being the 2-$\sigma$ ridge lines (i.e. 95\%
confidence interval).
%  Also plotted are the
%LMC Cepheids that have periods within the range for each chip, shifted to 
%minimise the least squares fit. 

The number of Cepheids used in the fits, the apparent distance moduli in 
$V$ and $I$ ($\mu_V, \mu_I$), reddenings $E(V-I)$, and extinction-corrected 
distance moduli $\mu_0$ for each photometry set are listed in 
Table~\ref{PL_fit_tab}.  The 1-$\sigma$\ uncertainties in $\mu_V$ and $\mu_I$ 
are internal errors, being the standard errors of the mean from the least squares fits 
(combined in quadrature with the uncertainties in the LMC PL fits).  Because 
a component of the scatter in the $V$ and $I$ PL relations will be correlated, 
the uncertainty in $\mu_0$ is derived from the dispersion in a Wesenheit 
function, $W = V - 2.45(V - I)$, which is adapted to the reddening assumed 
between $V$ and $I$, as this will implicitly propagate the photometric 
uncertainties and cancels the differential reddening-induced scatter 
(\markcite{mado82}Madore 1982).  The uncertainty in $\mu_0$ in 
Table~\ref{PL_fit_tab} thus includes the observational uncertainties in the 
photometry.  Ideally we would like to search for any possible 
zero-point variations between chips, but with so few Cepheids it would be 
impossible to tell if any differences were due to a 
zeropoint variation or simply a statistical fluctuation. For the record,
for chips 2 and 4, the $\mu_V$ were within 1.5-$\sigma$ of the all chip mean
for ALLFRAME and DoPHOT, but the chip 3 $\mu_V$ was 3-$\sigma$ larger for 
ALLFRAME and 4-$\sigma$ larger for DoPHOT.
%We note, however, that \markcite{hill98}Hill et al.\ (1998) found 
%similar discrepancies 
%with the WF3 photometry in the M100 data, but assumed this was due to the 
%extra crowding (for M100, WF3 was the closest chip to the nucleus).  For
%NGC~7331, WF3 is the least affected by crowding, implying the cause may be
%related to the WF3 chip and/or optics.  It is also possible that the difference
%is due to a calibration error in the WF3 magnitudes, but as these were made 
%independently for DoPHOT and ALLFRAME, such an error, if it exists, is more
%likely to be in the \markcite{holt95b}Holtzman et al.\ (1995b) calibration from 0.5\arcsec\
%apertures to standard magnitudes.  However, as there are only three
%Cepheids on WF3, if there is any systematic error involved, it will have
%minimal impact on the average derived from all three WF chips.
The luminosity function for stars with Cepheid-like colors in 
Figure~\ref{LF_fig} showed that we may be affected by incompleteness 
at the faintest (i.~e.\ short period) end of the Cepheid distribution.  
To check for the effects this may have in fitting the PL relations, the
lower period limit was also varied, from 10 days (the lower period limit)
to 15, 20, and 25 days (Table~\ref{PL_fit_tab}).  All fluctuations in
$\mu_0$ relative to the all-Cepheid fit are within the uncertainties,
and there is no strong systematic trend in $\mu_0$, as the low period limit 
is increased from 10 to 25 days.  This is due to our keeping the slope
of the PL relation fixed, and indicates that no strong Malmquist bias 
is present in our Cepheid sample.  Thus our distance estimate will not 
be significantly affected by incompleteness.  

The distance moduli derived from the all-Cepheid sample for ALLFRAME and 
DoPHOT agree within one sigma, indicating no significant difference between
the two procedures.  We note that the DoPHOT distance modulus for periods 
above 15 days agrees very well with the ALLFRAME result, as would be expected
given the higher signal to noise ratio of these brighter Cepheids.
For consistency with previous papers, and for simplicity, we henceforth quote 
the results based on the ALLFRAME photometry alone.

The reddening of $E(V-I) = 0.19$ (ALLFRAME), implies a mean 
extinction to NGC~7331 of $A_V = 0.47\pm0.20$ mag.\footnote{We thank the 
referee for pointing out that extinctions should formally be calculated in 
the observed passbands. However, for stars of Cepheid-like colors this makes 
very little difference.  Using the mean extinction values derived from Tables 
12a and 12b of \markcite{holt95b}Holtzman et al.\ (1995b) for E($B-V$) = 0.15 
(i.e. E($V-I$) = 0.19), then $A_{F555W} = 0.46$, almost identical 
to the value for $A_V$, and well within the observational uncertainties.}
The foreground Galactic extinction in the direction 
of NGC~7331 is $A_V = 0.27$ (derived from $E(B-V)=0.083$ from 
\markcite{burs84}Burstein \& Heiles 1984), thus the mean extinction internal 
to the NGC~7331 Cepheid field is 0.20 mag in $V$.  
The extinction-corrected distance modulus is 30.89 mag,
equivalent to a distance of 15.1 Mpc.

\subsection{Errors in the Distance Modulus}

The various contributions to the total error budget are listed in 
Table~\ref{Error_tab}.
In addition to the uncertainties of the photometry (included in the PL fits), 
the true distance estimate will also be affected by the systematic 
uncertainties in the zero-point calibration from HST to ground-based $V$ and 
$I$ magnitudes, and the LMC distance modulus.  
The former are assumed to be uncorrelated, and hence must be factored for 
the ratio of their contributions via the extinction correction: i.~e.\ 
$\mu_0 = 2.45\mu_I - 1.45\mu_V$, hence $\Delta(ZP) = ((2.45\Delta(ZP)_I)^2
+ (1.45\Delta(ZP)_V)^2)^\frac{1}{2}$.

Differences in metal abundance between our field in NGC~7331 and
the calibrating Cepheids in the LMC might affect our derived distance.
Both the empirical test in M31 performed by \markcite{free90b}Freedman \& 
Madore (1990), and the theoretical calculations of \markcite{chio93}Chiosi, 
Wood, \& Capitanio (1993), suggested that the magnitude of such a dependence 
is relatively weak.  On the other hand, \markcite{goul94}Gould (1994), 
using the same data as \markcite{free90b}Freedman \& Madore (1990), 
but employing a different method of analysis, claimed a sizable effect
of $\Delta \mu \sim 0.6 \Delta$[Fe/H], although there were inconsistencies
between different bands.  \markcite{stif95}Stift (1995)  also used the
\markcite{free90b}Freedman \& Madore (1990) data, in yet another analysis, 
and found a similarly sized effect as found by \markcite{goul94}Gould (1994).
Apart from the different methodologies used, much of the disparity between 
these three studies, based as they are on identical data (38 Cepheids in M31, 
spanning a range of metallicity $\Delta$[Fe/H] = 0.75), is due to the 
differing ways in which the reddening is measured (the effects of reddening 
and metallicity both act in the same color direction), and is partly due to 
the small number of Cepheids in the metal poor field (only eight), and to the 
fact that M31 has reasonably large amounts of reddening which must be 
corrected.  More recently, \markcite{sass97}Sasselov et al. (1997) find 
evidence for an intermediate metallicity dependence ($\Delta \mu \sim 0.4 
\Delta$[Fe/H]) between the LMC and SMC Cepheids.  Although this study enjoys 
a surplus of Cepheids (500), it also suffers from inherent uncertainties due 
to the smaller range in metallicity ($\Delta$[Fe/H] = 0.35), plus extra 
uncertainties due to the unknown distance between the LMC and SMC, and 
the considerable depth distribution for the SMC Cepheids.   To sidestep
most of these uncertainties, we have made an empirical test of 
the metallicity dependence of the PL relation, based on observations of 
two fields in M101 (\markcite{kenn98}Kennicutt et al.\ 1998).
These samples in M101 have the advantage of spanning a much larger range 
of metallicity ($\Delta$[Fe/H] = 1.4; \markcite{zari90}Zaritsky et al.\ 1990), 
while also having reduced reddening due to M101 being almost face-on.
The empirical metallicity dependence is $\Delta \mu = 0.24(\pm0.16)
\Delta$[O/H], in the sense that a target sample of Cepheids more metal rich 
than the LMC would appear to be closer than it actually is, and
$\Delta$[O/H] is the difference in [O/H] abundance, in dex, between the target 
galaxy and the LMC.

As argued in \markcite{kenn98}Kennicutt et al.\ (1998), a metallicity effect of
this magnitude is barely detectable in the metallicity range of our Key 
Project galaxies.  In particular, for NGC 7331, the \ion{H}{2} region 
abundances measured by \markcite{oey93}Oey \& Kennicutt 
(1993) and \markcite{zari94}Zaritsky, Kennicutt, \& Huchra (1994), and  
recently supplemented by our own measurements of an \ion{H}{2} region in the 
Cepheid field itself, indicate an oxygen abundance of about $1.5\pm0.3$ 
times that of the LMC, which would correspond to a correction of only 
$+0.05\pm0.04$ mag in distance modulus.    Given the small 
size of the effect and its large uncertainty, and in order to be consistent 
with previous galaxies in this series, we have not applied it to our data, 
but simply add this as a possible systematic error in Table~\ref{Error_tab}.
Once the Cepheid distances to all Key Project galaxies have been obtained, we
intend using the combined data set, with its large range in metallicity, to 
reduce the uncertainty in the metallicity effect, and apply it to all galaxies
prior to calibrating the secondary distance indicators. 

\subsection{Comparison with Previous Distance Estimates}

Previously published distance estimates to NGC~7331 are listed in 
Table~\ref{Dist_tab}.  \markcite{sers60}Sersic (1960) and 
\markcite{osma82}Osman (1982) used the size distributions of \ion{H}{2} regions
and dust clouds, respectively, to estimate distances, and 
\markcite{vand60}van~den~Bergh (1960) used a calibration of luminosity classes.
\markcite{rubi65}Rubin et al.\ (1965) estimated a 
velocity distance of 14.4 Mpc derived from a recession velocity of +1082 km/s 
(corrected only for Galactic rotation) and assuming H$_0 = 75$ km/s/Mpc.
They also noted the similarities between M31 and NGC~7331, and by assuming
similar diameters, made a further distance estimate in terms 
of the ratio of their apparent diameters (20:1).  This estimate converts 
to 15.4 Mpc, where we adopt the distance to M31 of 770 kpc from 
\markcite{free90b}Freedman \& Madore (1990).  
%Interestingly, if we also apply the same reasoning to their 
%luminosities (i.~e.\a one-dimensional Tully-Fisher relation), then their
%apparent $H_{-0.5}$ magnitudes of 0.91 for M31 and 6.44 for NGC~7331 
%(\markcite{aaro82a}Aaronson et al.\ 1982a) imply a distance to NGC~7331 of
%just 9.7 Mpc, implying that galactic masses are more closely tied to 
%diameters than luminosities.  
\markcite{balk73}Balkowski et al.\ (1973) used a variety of correlations
between estimators of mass, luminosity, morphology, and \ion{H}{1} density, 
to derive a mean distance estimate, with a large scatter, of $13^{+9}_{-5}$ Mpc.
\markcite{tull88}Tully (1988) estimated a velocity distance of 14.3 Mpc  
based on assuming H$_0 = 75$ km/s/Mpc, and a Virgo-centric infall model of 
the Local Group of 300 km/s, while the velocity data of 
\markcite{aaro82a}Aaronson et al.\ (1982a), when coupled with the Virgo model 
of \markcite{aaro82b}Aaronson et al.\ (1982b), imply a distance of 
12 Mpc.  The only explicit Tully-Fisher distance estimate for NGC~7331 is in 
the $B$\ band, and is 9.6 Mpc (\markcite{bott85}Bottinelli et al.\ 1985).  
However, \markcite{aaro82a}Aaronson et al.\ (1982a) published the IRTF 
parameters ($H_{-0.5}$ and $\Delta V(0)$) for NGC~7331, which converts to 
an IRTF distance of $10.1 ^{+2.0}_{-1.7}$ Mpc, using the calibration of 
\markcite{aaro80}Aaronson, Mould \& Huchra (1980) 
(i.e. $H^{abs}_{-0.5} = -21.23 - 10.0[$log$\Delta V(0) - 2.5], \sigma = 0.4$).
Using the same calibration, \markcite{aaro83}Aaronson \& Mould (1983) derived
a distance to the NGC 7320/7331 pair of $12.9 ^{+3.5}_{-2.8}$ Mpc.

The only recorded supernova in NGC~7331 was SN1959D, of type IIL, unfortunately
occurring before the era of estimating distances to type II supernovae from 
the expanding photospheres method (e.~g.\ \markcite{schm94}Schmidt et al.\ 
1994\footnote{Although \markcite{schm94}Schmidt et al.\ list SN1989L as 
occurring in NGC~7331, this should read NGC~7339.}). 

\section{Discussion}
\label{Disc}

The aim of the Key Project is the measurement of the Hubble Constant by
calibration of secondary distance indicators. We consider the impact of
our new distance on the calibration of two such distance indicators here.

\subsection{The Tully Fisher relation}

\markcite{moul96}Mould  et al.\ (1996) have updated the calibration of the 
infrared Tully Fisher (IRTF) relation, noting the importance of populating the
calibration for galaxies of large line width, as these are galaxies which 
can be detected at large distances, where peculiar velocities are negligible 
relative to recession velocities. In Table~\ref{TF_tab} and Figure~\ref{TF_fig}
we add NGC~7331 to the TF calibration derived from all suitable galaxies which 
now have Cepheid distances.  We use the $H_{g}$
magnitudes of \markcite{torm95}Tormen \& Burstein (1995), which are the
\markcite{arro82a}Aaronson et al.\ (1982a) $H_{-0.5}$ magnitudes re-calculated to
agree with the optical diameters in RC3, and which do not include any
corrections for inclination.  For any given galaxy, the major effect of these
changes is that the $H_{g}$ magnitudes are about 0.35 magnitudes
brighter than the  $H_{-0.5}$ magnitudes.  We adopt an \ion{H}{1} 
velocity width of $531 \pm 10$ km/s (given by \markcite{fish81}Fisher \& 
Tully), an inclination for NGC~7331 of 73\arcdeg, in agreement with 
\markcite{torm95}Tormen \& Burstein (1995), but 2\arcdeg\ less than that
of \markcite{aaro82a}Aaronson et al.\ (1982a). 

The least squares fit IRTF relation excluding NGC~7331 is
$H_{g} = -9.6(\pm2.4)({\rm log}\Delta {\rm V} - 2.5) - 21.7(\pm 0.3)$,
with $\sigma = 0.35$ (the dashed-dotted line in Figure~\ref{TF_fig}).
  When we add in NGC~7331 the fit becomes
$H_{g} = -10.2(\pm2.2)({\rm log}\Delta {\rm V} - 2.5) - 21.8(\pm0.3)$,
with $\sigma = 0.39$ (the solid line in Figure~\ref{TF_fig}).  
%When using the Aaronson H mags, the fits are:
%$H_{-0.5} = -9.9(\pm2.4)({\rm log}\Delta {\rm V} - 2.5) - 21.4(\pm 0.3)$,
%with $\sigma = 0.37$.  When we add in NGC~7331 the fit becomes
%$H_{-0.5} = -10.4(\pm2.2)({\rm log}\Delta {\rm V} - 2.5) - 21.4(\pm0.3)$,
%with $\sigma = 0.39$.  
Hence this distance to NGC~7331 is 
consistent with the current IRTF calibration, but it points to a dispersion 
approaching 0.4 mag in the relation, not very different from that found
by \markcite{aaro83}Aaronson \& Mould (1983) in groups and clusters. 
Such a dispersion is highlighted by the fact that although NGC~7331 and M31 
have very similar line widths and diameters, NGC~7331 is one magnitude brighter
in $H_{g}$, implying an $M/L$ lower by a factor of $\sim$2.5, possibly
due to a different star formation history.

% However, an interesting
%trend can be seen in Figure~\ref{TF_fig}, in that all but one (M96) of the 
%HST-measured distances are consistently producing $H_{-0.5}$ brighter by
%$\sim0.8$ magnitudes!!!????

The Tully-Fisher relation is a dynamical relation, and improved understanding
of the kinematics and mass to light ratio of the calibrator galaxies can
contribute to developing the relation.
The kinematics of NGC~7331 have been extensively analyzed by
\markcite{bowe93}Bower et al.\ (1993), who were able to fit a constant
$M/L_V$  dynamical model, as a function of radius, and found $M/L_V = 5.3$,
assuming a distance of 12 Mpc (from \markcite{aaro82a}Aaronson et al.\ 1982a).
Our new distance therefore implies an $M/L_V = 4.2$, and a total mass
of $4.6 \times 10^{11} M_{\sun}$.  This is comparable to a mass
derived from \markcite{fish81}Fisher \& Tully (1981) of $4.0 \pm 0.5 \times 
10^{11} M_{\sun}$, and $M/L_B = 7.9$ (where our
distance of 15.1 Mpc implies a Holmberg diameter $A_H = 65$ kpc,
$\mu_0 = 30.88$ implies $M_B = -21.38$ and log$L_B = 10.74$).

Although NGC~7331 looks like a normal early-intermediate type spiral galaxy
with a significant bulge, it is also claimed to have a mildly active (weak LINER) 
nucleus. \markcite{cowa94}Cowan et al.\ (1994) measured its 20 cm and 6 cm 
radio flux to be 234 and 121 $\mu$Jy, respectively.  At a distance of 
15.1 Mpc, this makes it 3.3 and 4.6 times the luminosity of Sgr A, at these 
wavelengths.  However, \markcite{ohya96}Ohyama \& Taniguchi (1996) suggest, 
on the basis of stellar population synthesis, that the apparent nuclear 
activity in NGC~7331 is due to a recent burst of star formation, rather than 
an active galactic nucleus.
 
%\subsection{The Virgo-centric flow}
%
%We can  derive a `naive' distance estimate based just on NGC~7331's recession 
%velocity.  Adopting \markcite{tull88}Tully's (1988) simple Virgo-centric 
%infall model of the Local Group, the Hubble flow velocity for NGC~7331 would 
%then be 1075 km/s, giving H$_0 = 70$ km/s/Mpc.  Unfortunately, such an 
%estimate is prone to systematic errors, by ignoring the likely extra 
%contribution to peculiar velocity provided by the rest of the mass outside 
%the local group and the Virgo cluster, which is why we prefer to use the 
%Cepheid distances as calibrators of secondary distance indicators capable 
%of measuring distances to redshiRTF_figfts where peculiar velocities are 
%comparatively small.  Alternatively, the Virgo-centric flow model of 
%Aaronson et al.\ (1982a) implies a distance 0.74 times that of the Virgo 
%cluster.  Our distance to NGC~7331 would then imply a kinematic distance 
%to Virgo of 20.7 Mpc, compared to a direct Cepheid distance (derived from 
%M100) of $16.1 \pm 3.2$\ Mpc (Ferrarese et al.\ 1996).  

\subsection{Surface Brightness Fluctuations}

In addition to the TF calibration, the large bulge in NGC~7331 can be
used as a calibrator for the surface brightness fluctuations (SBF) method.  
\markcite{tonr97}Tonry et al.\ (1997) have adopted a SBF  calibration of 

\begin{equation}
\overline{M_I} = \overline{M_{I0}} + 4.5 ~( V - I - 1.15)
\end{equation}

At a distance of 15.1 Mpc, their measurement of a field 1-2 arcmin
along the minor axis of NGC~7331 yields $\overline{M_{I0}} = -2.3 \pm 0.3$ mag,
significantly brighter than their adopted value of $-1.74\pm 0.07$ mag 
derived from the mean calibration (\markcite{tonr97}Tonry et al.\ 1997).
The distance to NGC 7331 from the Cepheids and SBF methods thus disagree
by 0.55 magnitude.  At least part of this disagreement is probably due to 
internal reddening in the NGC~7331 bulge field.  Although Tonry et al.\ 
take into account the foreground reddening to each of the SBF samples 
(e.~g.\ A$_B = 0.33$ mag for NGC~7331), no correction is made for any 
internal extinction.  
For the E and S0 galaxies, and the very edge-on spirals, this may be 
appropriate, but for NGC~7331 there is evidence from stellar population studies
in the bulge by \markcite{prad96}Prada (1996) that there is additional 
reddening internal to the bulge, implying an extinction of A$_V \sim 0.5$ mag, 
or A$_I = 0.3$ mag.  A quirk of SBF is that taking reddening into account 
causes $\overline{M_{I0}}$ to become fainter.  Hence an extra reddening of 
A$_I=0.3$ would increase the value of $\overline{M_{I0}}$ quoted here, making 
it more consistent with Tonry's standard value.  As well, if there is this 
much dust in the bulge of NGC~7331, then it may also have a young stellar 
component as well.  Although the color term in the SBF calibration is intended 
to allow for such variations, the reddening would tend to mask the bluer 
colors such a young component would contribute, thus making the deviation from 
an unreddened old population even greater.  These are only possibilities, of 
course.  A larger 
sample of SBF calibrators will allow us to search for hidden variables in the 
calibration, just as it does for Tully-Fisher.

\subsection{The NGC~7331 Group}

According to \markcite{garc93}Garcia (1993), NGC~7331 is the dominant member 
of a group with three other galaxies:  UGC12082, NGC~7320A, and UGC12060, 
which, relative to NGC~7331, are all within a radius of 1.7\arcdeg\ on the 
sky and within $62$ km/s in radial velocities.  Garcia also claimed NGC~7363 
to be a member of this group, as its heliocentric velocity is 830 km/s in RC3, 
but \markcite{wegn93}Wegner, Haynes \& Giovanelli (1993) measure its \ion{H}{1}
velocity as 6722 km/s, and thus a member of the Pisces-Perseus ridge or 
super cluster (along with various other galaxies which are nearby, on the sky, 
such as NGC~7320B, NGC~7320C, NGC~7335, NGC~7337, NGC~7369, and NGC~7340).   
Our group selection is based on a compilation of all galaxies within a radius 
of 9 degrees of NGC~7331 (at the distance of NGC~7331, 9 degrees corresponds 
to 2.4 Mpc), derived from a combination of the CfA Redshift Catalog 
(\markcite{huch97}Huchra 1997) and the NASA Extragalactic Database.  They are 
listed in Table~\ref{Group}, which gives their name, position (RA and Dec at 
J2000), galactocentric velocity, and radius on the sky (in degrees, relative 
to NGC~7331).  The group has a velocity dispersion of 64 km s$^{-1}$ (it might 
be argued that NGC~7292 and NGC~7217 should be excluded, as the remaining 
galaxies would then have a velocity dispersion of only 44 km s$^{-1}$).  
NGC~7331 is the brightest galaxy in this small group of 8-10 members.  
The second brightest galaxy is NGC~7217, about 6 degrees away.  At 15 Mpc, 
6 degrees is only about 1.5 Mpc. NGC~7457 is the third brightest member, 
and almost all the other galaxies in the group are more than three magnitudes 
fainter than NGC~7331.  This is a moderately isolated group, almost opposite 
Virgo, and almost 90 degrees away from Fornax, hence its well determined mean 
velocity of $1110 \pm 20$\ km s$^{-1}$\ (or $1088 \pm 16$\ km s$^{-1}$\ 
without NGC~7292 and NGC~7217) should provide a stable point of reference 
in studying local velocity flows.

\section{Conclusion}
\label{Conc}

We have used 15 epochs of HST F555W frames to produce light curves of all stars
within the field, to a limiting magnitude of $V \sim $28 mag, using two 
photometry programs (DoPHOT and DAOPHOT/ALLFRAME), and from these
light curves identified a total of 13 reliable Cepheids in NGC~7331.  Using 
the ALLFRAME data, and an additional 4 epochs of F814W frames, 
period-luminosity relations derived from LMC Cepheids were fitted to the 
NGC~7331 Cepheids, to measure a mean reddening of $E(V-I) = 0.18\pm0.06$, 
and an extinction-corrected distance modulus of  $30.89$ mag (uncertain
to $\pm0.16$ mag (random) and $+0.05\pm0.13$ mag (systematic)), equivalent 
to a distance of $15.1 ^{+1.4}_{-1.3}$ Mpc.

\acknowledgments

We are indebted to the staff at STScI for their very able support,
in particular to Doug Van Orsow.  We also thank John Tonry for early
access to the SBF results, and the referee for the helpful comments and
suggestions which have clarified many parts of the paper.
Support for  this  work  was  provided  by NASA through  grant  number
2227-87A from STScI which  is operated by  AURA under NASA  Contract 
NAS5-26555, and by a NATO Collaborative Research Grant CRG960178. This 
research has made use of the NASA/IPAC Extragalactic Database 
(NED) which is operated by JPL, California Institute of Technology, under 
contract with NASA.

\newpage

\appendix

\section{Appendix A - Calibration Parameters}

There are several differences between DoPHOT and DAOPHOT/ALLFRAME in the way 
the frames are prepared and PSF fitting is done for each filter set.  In 
DoPHOT the frame pairs are first combined to remove cosmic rays; the width 
and elongation of an analytic PSF is determined for each frame, but does not 
vary across a frame; this is compensated by measuring aperture corrections 
which vary across the frame, but do not vary from frame to frame.  In contrast,
and as the name implies, ALLFRAME measures the PSF magnitude of all frames;
the PSF fitted is a combination of an analytic function and a residual map,
both of which are allowed to vary across each frame; and aperture corrections
do not vary across each frame (they are also allowed to vary from frame to 
frame if there are enough isolated stars, but this was not the case for 
NGC~7331). As well as these differences in measuring instrumental photometry, 
there are also differences in calibration.

The calibration of the raw DoPHOT photometry is described in detail in
\markcite{saha96}Saha et al.\ (1996).  This calibration procedure has been
applied systematically to all DoPHOT photometry in this Key Project series,
and is given by:

\begin{equation}
{\rm HST}_{0.5} = {\rm DoPHOT}_{\rm obs} + 30.0 + AC + 2.5*{\rm log}_{10}(t) + ZP
\end{equation}

\noindent where HST$_{0.5}$ mag is the 0\arcsec.5 aperture magnitude system 
(F555W$_{0.5}$ and F814W$_{0.5}$), as defined by \markcite{holt95b}Holtzman 
et al.\ (1995b), AC is the aperture correction (to correct for the 9$\times$9 
pixel square aperture, derived from observations of Leo~I), $t$ is the 
exposure time, in seconds, and ZP is the zero-point, derived from comparing 
magnitudes obtained from the first epoch NGC~7331 frame with observations
of Pal~4, following the prescription of \markcite{holt95b}Holtzman et al.\ 
(1995b).  

The DoPHOT aperture corrections are given by:

\begin{equation}
AC = C0 + C1*x + C2*y + C3*x*x + C4*y*y + C5*x*y
\end{equation}

\noindent where x and y are the pixel coordinates of each full ($800\times800$)
WFPC2 chip, and the C coefficients are given in Table~\ref{App_AC_doph}
(the C0 coefficients also contain an additional offset derived by measuring
a mean aperture correction from the first epoch of the NGC~7331 frames).

As the calibration to HST$_{0.5}$ mag is done via cluster data, which have a 
faint background, these were corrected for the CTE effect, by adding a mean 
correction of $-0.02$ mag, which is an approximation to the recommended 
correction in \markcite{holt95b}Holtzman et al.\ (1995b) (which is to apply a 
ramp of $-0.04 {\rm mag}/800$\  Y pixels).  As discussed in 
\markcite{hill98}Hill et al.\ (1998),
zero-points for WFPC2 also seem to differ by 0.05 mag between
short and long exposures (i.~e.\ objects observed with long exposure times
appear brighter by 0.05 mag, compared to short exposure times).
For long exposures (i.~e.\ ours) the DoPHOT zero-points (for a gain of 7) 
are defined in Table~\ref{App_ZP_AC}.

Because DoPHOT and DAOPHOT have different methods of estimating the sky
and fit different PSF profiles, their calibration to standard magnitudes has
to be done independently.  To gain full advantage of this effort, the
ALLFRAME calibration was also done using a different data set, based on
exposures of stars in Galactic globular clusters, to provide an independent 
check on the calibration procedure.  Because ALLFRAME allows the PSF to vary 
across each chip, its calibration only requires zero-points and a single 
aperture correction per chip/filter to convert the raw magnitudes to 
equivalent 0.5 arcsec aperture magnitudes, where for DAOPHOT/ALLFRAME:

\begin{equation}
{\rm HST}_{0.5} = {\rm ALLFRAME}_{\rm obs} + AC + 2.5*{\rm log}_{10}(t) + ZP
\end{equation}

The zero-points were derived for each chip and filter from weighted averages 
of long exposure HST vs ground-based photometry of the clusters NGC~2419 and 
Pal~4 (Bolte \& Stetson, private communication).  A mean CTE ramp
effect of 0.040 mag/800 pixels was measured, and was then applied to the 
input cluster magnitudes.  Mean aperture corrections for each chip/filter
combination were derived from curve of growth models derived from fits to
isolated stars in each of the Key Project galaxies.  These, and the zero-points 
for ALLFRAME (again, on the LONG exposure scale) are given in Table~\ref{App_ZP_AC}.

The HST magnitudes were then transformed to 
Johnson $V$ or Kron-Cousins $I$ magnitudes, using the transformations 
derived by \markcite{holt95b}Holtzman et al.\ (1995b):

\begin{equation}
V = {\rm F555W}_{0.5} - 0.045(V - I) + 0.027(V-I)^2
\end{equation}

\begin{equation}
I = {\rm F814W}_{0.5} - 0.067(V - I) + 0.025(V-I)^2
\end{equation}

\noindent where the $(V - I)$ colors were estimated from iteration, starting
from F555W$_{0.5} - $F814W$_{0.5}$, and ending when 
$|V - I - (V - I)| < 0.001$ mag (usually one iteration was sufficient).

\clearpage

{\bf Figure Captions}

\figcaption[n7331_f01.eps]{A $9.5\times 9.5$ arcmin $V$\ image of NGC~7331, 
obtained with the Prime Focus TEK camera on the 
Isaac Newton Telescope, La Palma.
The position of the WFPC2 field is indicated by the chevron, and the
position of SN1959D is also marked.  North is to the RIGHT, and East is UP
(for purely aesthetic reasons).  The background galaxies to 
the east are (from south to north) NGC~7337, NGC~7335, and NGC~7336.
\label{INT_fig}}

\figcaption[n7331_f02.eps]{An F555W medianed image of the WFPC2 mosaic 
(with North UP, East to the LEFT).  The positions of the 
Cepheids in Table~\ref{good_ceph_tab} are marked by labeled circles.
\label{wfpc2_cephs}}

\figcaption[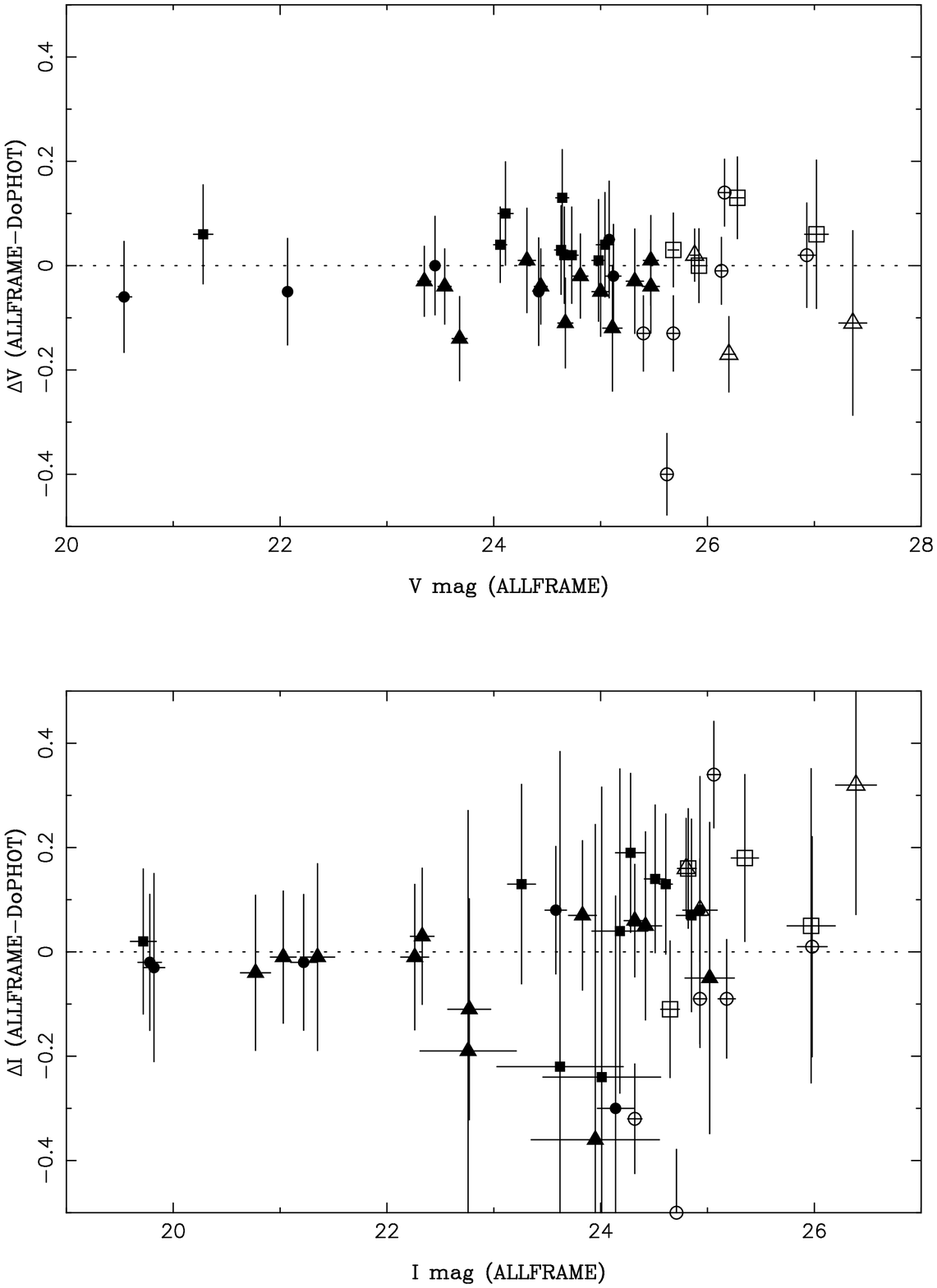]{Comparison of calibrated $V$ 
and $I$ photometry of the reference stars from chips 2-4 in 
Table~\ref{Ref_mags_tab} (filled symbols) and the Cepheids in 
Table~\ref{good_ceph_tab} (open symbols).  
Circles, triangles, and squares are stars from chips 2, 3, and 4, respectively.
\label{doC_alf_comp_fig}}

\figcaption[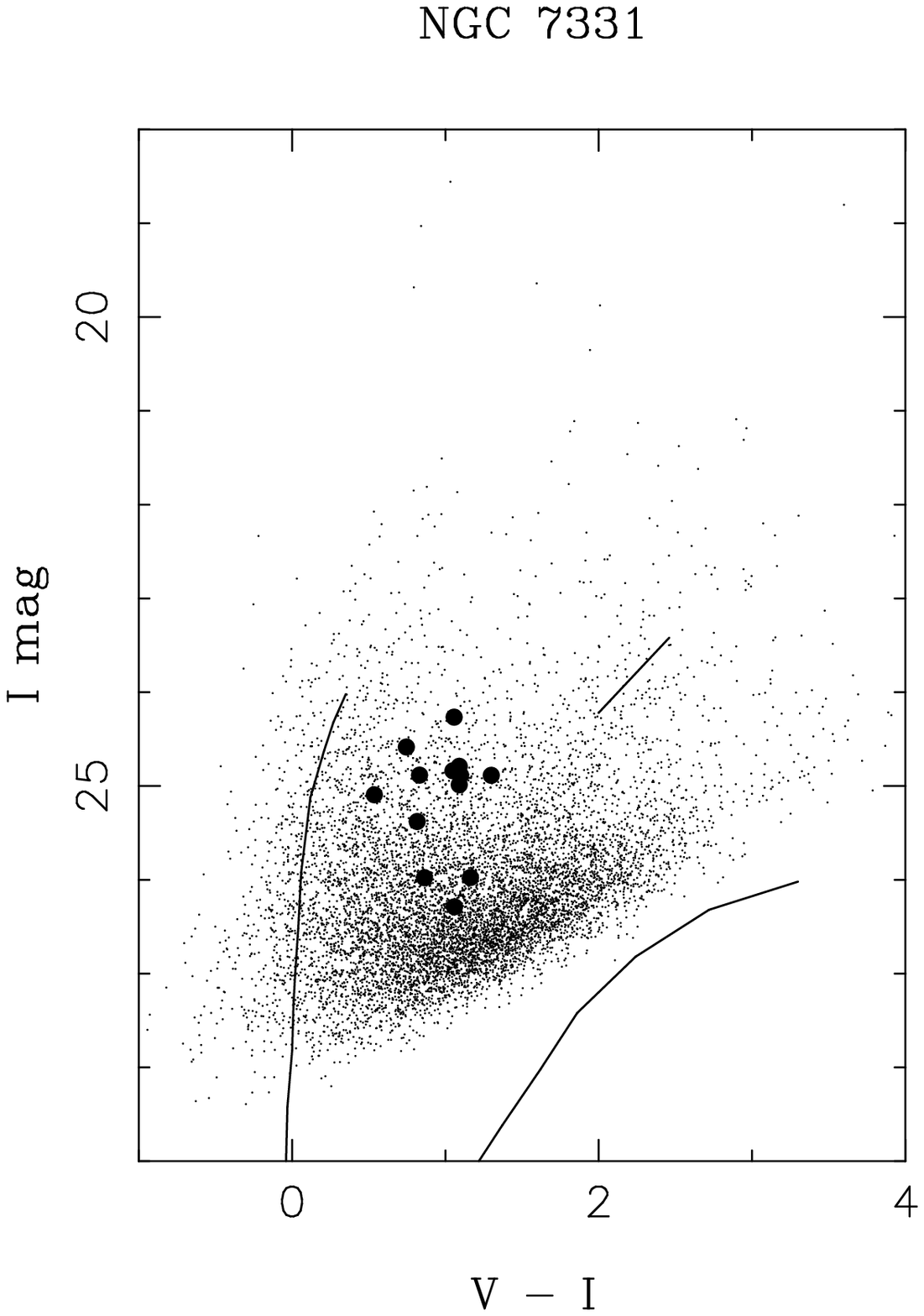]{$I$ vs $V - I$ color-magnitude diagram for
all chips, using the calibrated ALLFRAME photometry (for clarity, only every 
tenth star is plotted).  Also shown
are the Cepheids from Table~\ref{good_ceph_tab} (filled circles), which all lie
within the instability strip.  The three lines represent the locus of
the blue plume and red giant and supergiant branches of Shapley Constellation III in the LMC,
shifted to the distance and reddening of NGC~7331.
\label{CMD_fig}}

\figcaption[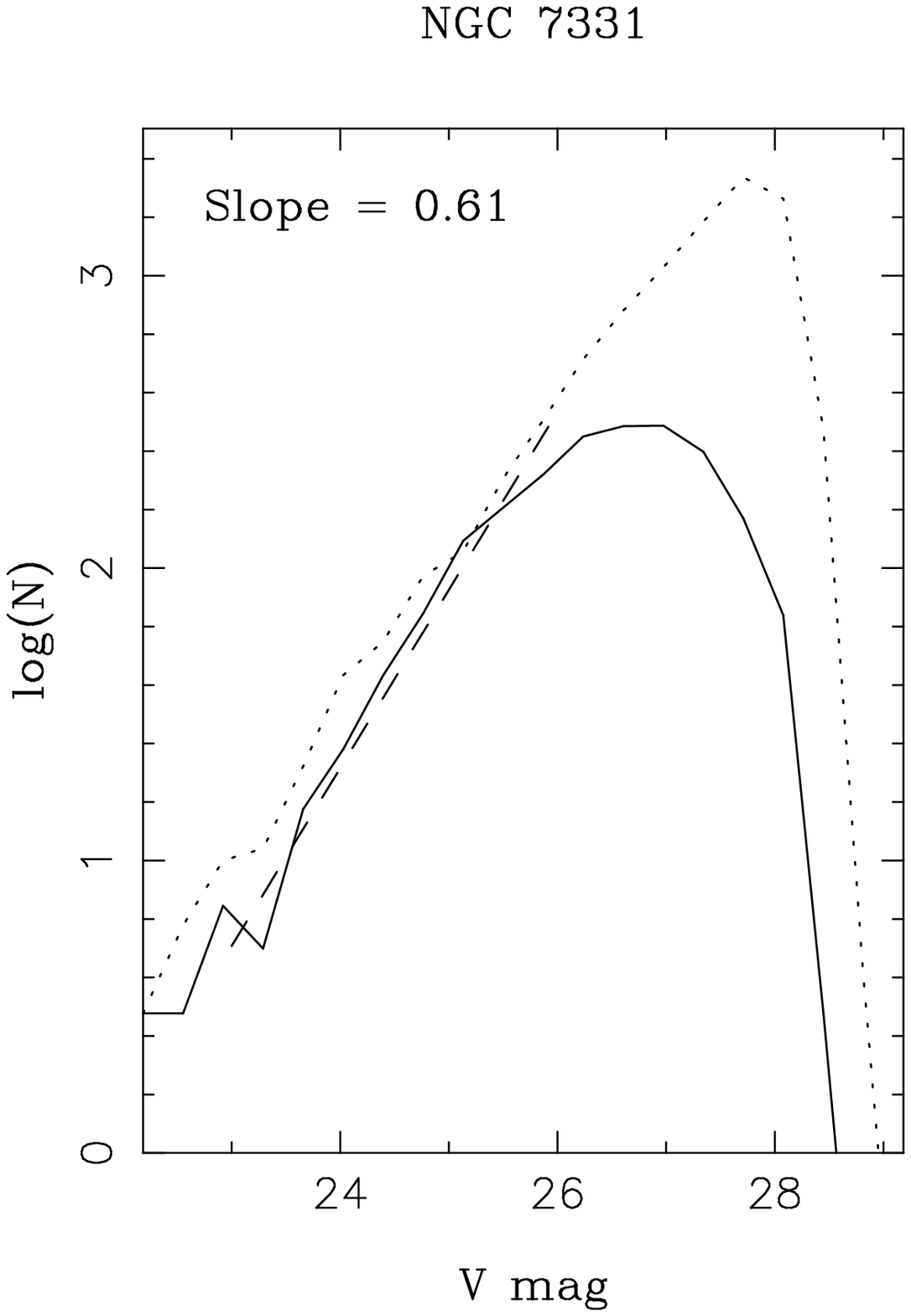]{The solid curve is the $V$ luminosity function 
for all stars with $V - I < $0.3 (ALLFRAME magnitudes), representing the blue 
plume.
The dashed line is a least squares fit to this luminosity function between
$V = 23$ and $V = 26$ mag.  The dotted curve is the $V$ luminosity function 
for all stars with $0.5 < V - I < $1.5 (ALLFRAME magnitudes).
\label{LF_fig}}

\figcaption[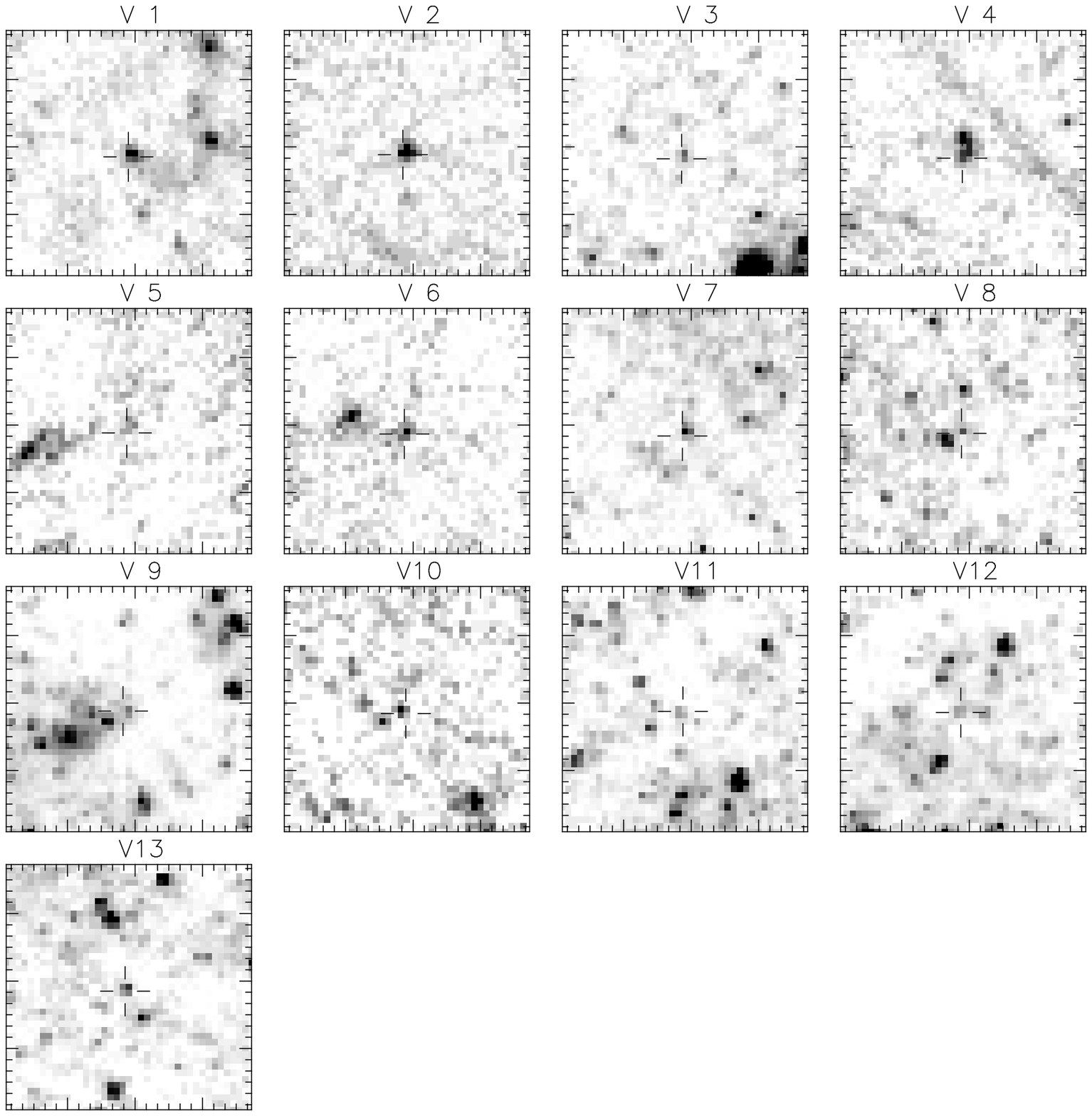]{Finding charts for each Cepheid listed in 
Table~\ref{good_ceph_tab}.  Each chart is centered on the Cepheid, and is
$4\times4$ arcsec squared (i.~e.\ $40\times40$ WF pixels).  North is up, east 
to the left.
\label{FC_fig}}

\figcaption[n7331_f07.eps]{Light curves, phased to the relevant 
period, for each Cepheid listed in Table~\ref{good_ceph_tab}.  On the left 
are the calibrated DoPHOT magnitudes, on the right are the calibrated ALLFRAME 
magnitudes.
\label{lc_fig}}

\figcaption[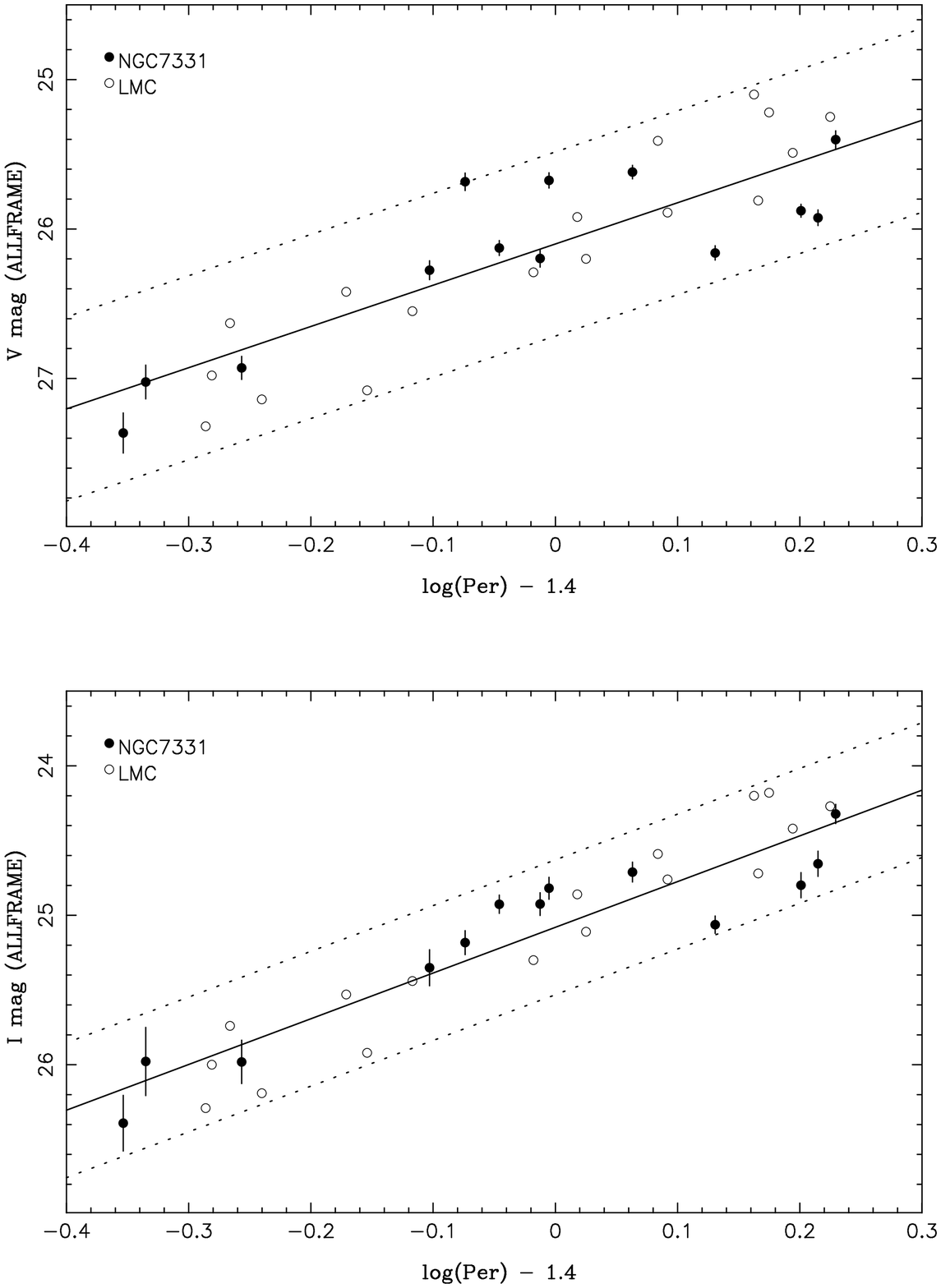]{$V$ (upper panel) and $I$ (lower panel) PL relations 
for the Cepheids listed in Table~\ref{good_ceph_tab} (calibrated ALLFRAME 
magnitudes).
The NGC~7331 Cepheids are the solid circles, and the LMC Cepheids are the
open circles.  The solid lines are least squares fits with the MF91 fixed 
slopes, and the dotted lines are the 95\% confidence interval (i.e.
the 2-$\sigma$ ridge lines).  
\label{PL_figH}}

\figcaption[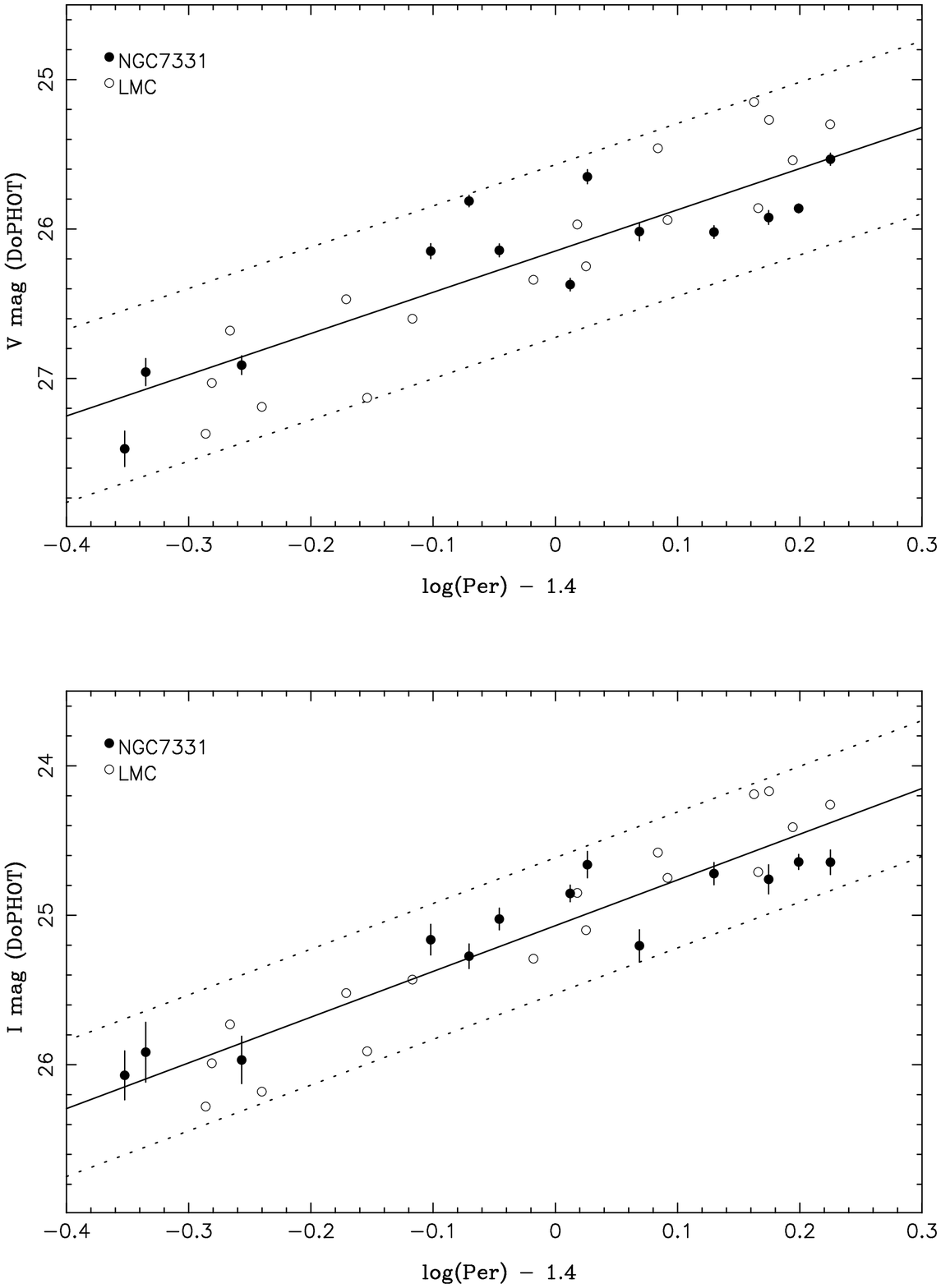]{$V$ (upper panel) and $I$ (lower panel) PL relations 
for the Cepheids listed in Table~\ref{good_ceph_tab} (calibrated DoPHOT 
magnitudes).  Symbols and lines as per Figure~\ref{PL_figH}.
\label{PL_figD}}

\figcaption[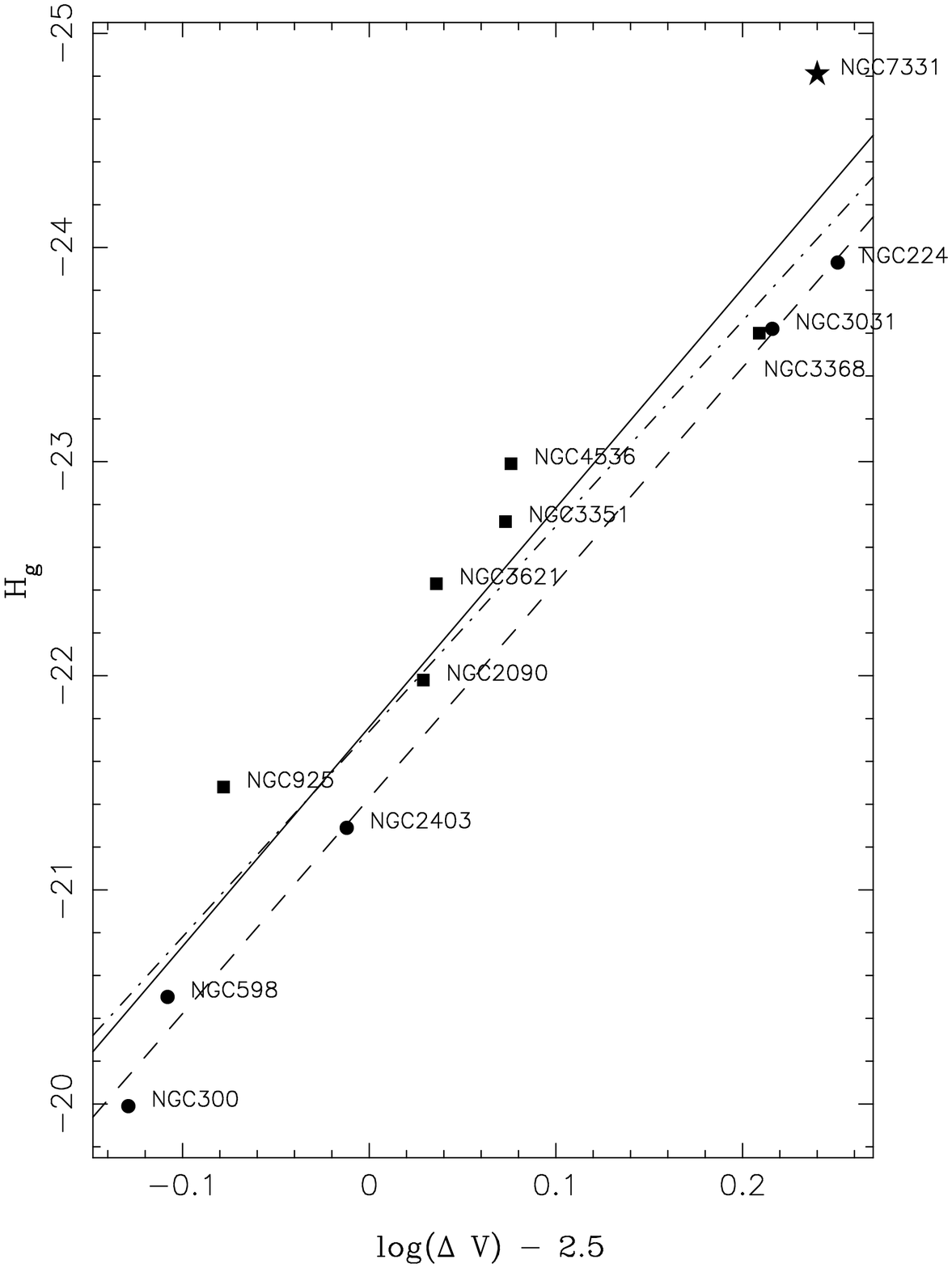]{IRTF absolute calibration, with absolute
$H_{g}$ magnitudes vs log($\Delta$V) - 2.5, from data in Table~\ref{TF_tab}.
Filled circles are the galaxies used by Freedman (1990), 
filled squares are the galaxies with recent
HST-derived Cepheid distances, and the filled star is NGC~7331.
The dashed line is a least squares fit solution to the Freedman 
(1990) set of galaxies, the dashed-dotted line is a least squares fit solution 
to all but NGC~7331, and the solid line is the least squares solution including
NGC~7331 (see text).
\label{TF_fig}}

\clearpage

\begin{deluxetable}{rclrccr}
\tablecaption{Epochs of Observations. \label{dates_tab}}
\tablewidth{0pt}
\tablehead{
\colhead{Epoch} & \colhead{Julian Date} & \colhead{Filename} & 
\colhead{Date} & \multicolumn{2}{c}{Exposure times (sec)} & \colhead{Filter} }
\startdata
 1 &  2449521.832 & u2781q01t & 18/06/94 & 1200 &  1600 &  F555W \nl
 2 &  2449530.612 & u2781r01t & 27/06/94 & 1200 &  1600 &  F555W \nl
 3 &  2449543.952 & u2781t01t & 10/07/94 & 1200 &  1600 &  F555W \nl
 4 &  2449545.762 & u2781u01t & 12/07/94 & 1200 &  1600 &  F555W \nl
 5 &  2449549.586 & u2781v01t & 16/07/94 & 1200 &  1600 &  F555W \nl
 6 &  2449552.806 & u2781w01t & 19/07/94 & 1200 &  1600 &  F555W \nl
 7 &  2449557.044 & u2781x02t & 24/07/94 & 1200 & \nodata &  F555W \nl
 8 &  2449561.927 & u2781y01t & 28/07/94 & 1200 &  1600 &  F555W \nl
 9 &  2449573.584 & u2781s01t &  9/08/94 & 1200 &  1600 &  F555W \nl
10 &  2449573.861 & u2782001t &  9/08/94 & 1200 &  1600 &  F555W \nl
11 &  2449581.226 & u2782101t & 17/08/94 & 1200 &  1600 &  F555W \nl
12 &  2449887.615 & u2o30101t & 19/06/95 & 1200 &  1600 &  F555W \nl
13 &  2449902.628 & u2o30201t &  4/07/95 & 1200 &  1600 &  F555W \nl
14 &  2449921.593 & u2o30301t & 23/07/95 & 1200 &  1600 &  F555W \nl
15 &  2449946.395 & u2o30401t & 17/08/95 & 1200 &  1600 &  F555W \nl
 1 &  2449521.966 & u2781q03t & 18/06/94 & 1200 &  1600 &  F814W \nl
 2 &  2449530.746 & u2781r03t & 27/06/94 & 1200 &  1600 &  F814W \nl
 9 &  2449567.754 & u2781z06t &  3/08/94 & 1200 & \nodata &  F814W \nl
12 &  2449887.631 & u2o30103t & 19/06/95 & 1200 &  1600 &  F814W \nl
\nl
13 &  2449902.645 & u2o30203t &  4/07/95 & ~260 & \nodata &  F555W \nl
13 &  2449902.650 & u2o30204t &  4/07/95 & ~260 & \nodata &  F814W \nl
\nl 
\enddata
\end{deluxetable}

\clearpage

%Produced by alf_do_cal_new.f
\begin{deluxetable}{lcrrrrrrrr}
\tablecaption{Reference Star Photometry. \label{Ref_mags_tab}}
\tablewidth{0pt}
\tablehead{
\colhead{ID} & \colhead{Chip} & \multicolumn{2}{c}{RA ~~ (J2000) ~~ Dec} & 
\colhead{X} & \colhead{Y} & 
\colhead{$V_{Alf}$} & \colhead{$I_{Alf}$} & 
\colhead{$V_{Doph}$} & \colhead{$I_{Doph}$} }
\startdata
 R  1 & 1 &  22:37:01.47   & +34:28:10.2   &   206.2 &   100.1 & $ 24.14 \pm 0.08 $ & $ 23.83 \pm 0.23 $ & $ 24.43 \pm 0.09 $ & $ 23.62 \pm 0.61 $ \nl
 R  2 & 1 &  22:37:00.40   & +34:28:09.0   &   211.7 &   394.2 & $ 21.91 \pm 0.03 $ & $ 20.26 \pm 0.04 $ & $ 22.06 \pm 0.08 $ & $ 20.05 \pm 0.38 $ \nl
 R  3 & 1 &  22:36:59.90   & +34:28:00.7   &   385.4 &   544.3 & $ 24.67 \pm 0.07 $ & $ 24.36 \pm 0.19 $ & $ 24.74 \pm 0.07 $ & $ 24.32 \pm 0.97 $ \nl
 R  4 & 1 &  22:36:59.69   & +34:28:04.2   &   301.9 &   595.6 & $ 25.47 \pm 0.09 $ & $ 22.36 \pm 0.12 $ & $ 25.53 \pm 0.06 $ & $ 22.38 \pm 0.53 $ \nl
\nl
 R  5 & 2 &  22:36:59.05   & +34:28:31.7   &   356.0 &   190.0 & $ 24.42 \pm 0.05 $ & $ 24.14 \pm 0.17 $ & $ 24.47 \pm 0.09 $ & $ 24.44 \pm 0.37 $ \nl
 R  6 & 2 &  22:36:56.71   & +34:28:31.9   &   648.0 &   214.9 & $ 22.07 \pm 0.02 $ & $ 19.82 \pm 0.10 $ & $ 22.12 \pm 0.10 $ & $ 19.85 \pm 0.15 $ \nl
 R  7 & 2 &  22:36:58.07   & +34:28:56.4   &   457.9 &   448.6 & $ 23.45 \pm 0.05 $ & $ 21.22 \pm 0.12 $ & $ 23.45 \pm 0.08 $ & $ 21.24 \pm 0.05 $ \nl
 R  8 & 2 &  22:36:57.52   & +34:28:56.0   &   525.8 &   450.4 & $ 25.08 \pm 0.05 $ & $ 23.58 \pm 0.10 $ & $ 25.03 \pm 0.10 $ & $ 23.50 \pm 0.07 $ \nl
 R  9 & 2 &  22:36:58.05   & +34:28:58.9   &   458.1 &   473.5 & $ 20.54 \pm 0.07 $ & $ 19.78 \pm 0.11 $ & $ 20.60 \pm 0.08 $ & $ 19.80 \pm 0.07 $ \nl
 R 10 & 2 &  22:36:57.94   & +34:29:05.9   &   465.7 &   545.4 & $ 25.12 \pm 0.07 $ & $ 24.93 \pm 0.16 $ & $ 25.14 \pm 0.07 $ & $ 24.85 \pm 0.20 $ \nl
\nl
 R 11 & 3 &  22:37:02.11   & +34:28:27.4   &   121.6 &   117.1 & $ 25.47 \pm 0.08 $ & $ 25.02 \pm 0.23 $ & $ 25.51 \pm 0.04 $ & $ 25.07 \pm 0.19 $ \nl
 R 12 & 3 &  22:37:01.90   & +34:29:13.0   &   582.4 &   127.2 & $ 23.54 \pm 0.06 $ & $ 22.76 \pm 0.45 $ & $ 23.58 \pm 0.04 $ & $ 22.95 \pm 0.10 $ \nl
 R 13 & 3 &  22:37:02.37   & +34:29:13.5   &   583.3 &   185.9 & $ 24.67 \pm 0.07 $ & $ 24.42 \pm 0.15 $ & $ 24.78 \pm 0.05 $ & $ 24.37 \pm 0.10 $ \nl
 R 14 & 3 &  22:37:02.48   & +34:29:03.8   &   484.5 &   192.9 & $ 24.81 \pm 0.07 $ & $ 23.83 \pm 0.13 $ & $ 24.83 \pm 0.04 $ & $ 23.76 \pm 0.06 $ \nl
 R 15 & 3 &  22:37:02.84   & +34:28:22.6   &    64.2 &   204.1 & $ 25.11 \pm 0.09 $ & $ 24.32 \pm 0.10 $ & $ 25.23 \pm 0.08 $ & $ 24.26 \pm 0.04 $ \nl
 R 16 & 3 &  22:37:03.06   & +34:28:36.1   &   200.0 &   243.1 & $ 24.44 \pm 0.06 $ & $ 23.95 \pm 0.60 $ & $ 24.48 \pm 0.04 $ & $ 24.31 \pm 0.07 $ \nl
 R 17 & 3 &  22:37:05.45   & +34:28:51.0   &   326.9 &   552.6 & $ 24.31 \pm 0.08 $ & $ 21.35 \pm 0.16 $ & $ 24.30 \pm 0.06 $ & $ 21.36 \pm 0.08 $ \nl
 R 18 & 3 &  22:37:05.57   & +34:28:49.7   &   312.6 &   567.1 & $ 25.47 \pm 0.07 $ & $ 22.77 \pm 0.20 $ & $ 25.46 \pm 0.05 $ & $ 22.88 \pm 0.07 $ \nl
 R 19 & 3 &  22:37:05.97   & +34:29:15.4   &   567.3 &   637.6 & $ 23.68 \pm 0.07 $ & $ 20.77 \pm 0.14 $ & $ 23.82 \pm 0.04 $ & $ 20.81 \pm 0.05 $ \nl
 R 20 & 3 &  22:37:06.12   & +34:29:02.0   &   430.4 &   645.0 & $ 23.35 \pm 0.06 $ & $ 21.03 \pm 0.12 $ & $ 23.38 \pm 0.03 $ & $ 21.04 \pm 0.04 $ \nl
 R 21 & 3 &  22:37:06.41   & +34:28:50.3   &   310.0 &   672.1 & $ 25.32 \pm 0.08 $ & $ 22.26 \pm 0.13 $ & $ 25.35 \pm 0.06 $ & $ 22.27 \pm 0.05 $ \nl
 R 22 & 3 &  22:37:06.45   & +34:29:13.6   &   545.3 &   695.1 & $ 25.00 \pm 0.08 $ & $ 22.33 \pm 0.11 $ & $ 25.05 \pm 0.03 $ & $ 22.30 \pm 0.07 $ \nl
\nl
 R 23 & 4 &  22:37:03.47   & +34:28:11.4   &   271.2 &   128.6 & $ 24.98 \pm 0.06 $ & $ 24.85 \pm 0.14 $ & $ 24.97 \pm 0.10 $ & $ 24.78 \pm 0.12 $ \nl
 R 24 & 4 &  22:37:02.85   & +34:28:03.3   &   189.8 &   207.1 & $ 24.73 \pm 0.06 $ & $ 24.51 \pm 0.10 $ & $ 24.71 \pm 0.07 $ & $ 24.37 \pm 0.10 $ \nl
 R 25 & 4 &  22:37:03.68   & +34:27:59.6   &   288.5 &   248.4 & $ 25.04 \pm 0.08 $ & $ 24.61 \pm 0.06 $ & $ 25.00 \pm 0.06 $ & $ 24.48 \pm 0.12 $ \nl
 R 26 & 4 &  22:37:04.11   & +34:27:50.9   &   339.3 &   343.2 & $ 21.28 \pm 0.09 $ & $ 19.72 \pm 0.12 $ & $ 21.22 \pm 0.03 $ & $ 19.70 \pm 0.07 $ \nl
 R 27 & 4 &  22:37:05.62   & +34:27:50.7   &   526.8 &   358.9 & $ 24.64 \pm 0.06 $ & $ 24.28 \pm 0.14 $ & $ 24.51 \pm 0.07 $ & $ 24.09 \pm 0.06 $ \nl
 R 28 & 4 &  22:37:05.09   & +34:27:48.6   &   459.0 &   374.7 & $ 24.63 \pm 0.06 $ & $ 24.18 \pm 0.26 $ & $ 24.60 \pm 0.06 $ & $ 24.14 \pm 0.17 $ \nl
 R 29 & 4 &  22:37:05.59   & +34:27:40.4   &   515.9 &   462.1 & $ 24.66 \pm 0.07 $ & $ 24.01 \pm 0.55 $ & $ 24.64 \pm 0.06 $ & $ 24.25 \pm 0.08 $ \nl
 R 30 & 4 &  22:37:05.21   & +34:27:33.8   &   463.8 &   525.6 & $ 24.06 \pm 0.06 $ & $ 23.62 \pm 0.59 $ & $ 24.02 \pm 0.04 $ & $ 23.84 \pm 0.13 $ \nl
 R 31 & 4 &  22:37:03.26   & +34:27:23.1   &   213.0 &   616.7 & $ 24.11 \pm 0.07 $ & $ 23.26 \pm 0.13 $ & $ 24.01 \pm 0.07 $ & $ 23.13 \pm 0.14 $ \nl
\nl
\enddata
\end{deluxetable}

\clearpage

\begin{deluxetable}{crrr}
\tablecaption{ALLFRAME $-$ DoPHOT Comparison. \label{Mags_comp_tab}}
\tablewidth{0pt}
\tablehead{
\colhead{Chip} & N & \colhead{$\Delta V$} & 
\colhead{$\Delta I$} }
\startdata
  1  &   4 & $-0.143 \pm 0.061 $ & $ +0.110 \pm 0.068 $ \nl
  2  &   6 & $-0.022 \pm 0.019 $ & $ -0.035 \pm 0.062 $ \nl   
  3  &  12 & $-0.050 \pm 0.015 $ & $ -0.047 \pm 0.037 $ \nl  
  4  &   9 & $+0.050 \pm 0.014 $ & $ +0.029 \pm 0.052 $ \nl  
\nl
\enddata
\end{deluxetable}

\clearpage

%Produced by n7331_ceph_plot_pap.f
\begin{deluxetable}{rccccc}
\tablecaption{Calibrated DoPHOT F555W$_{0.5}$ and F814W$_{0.5}$ magnitudes of 
Cepheids with good light curves. 
\label{good_cephotD}}
\tablewidth{16cm}
\tableheadfrac{0.0}
\tablehead{Epoch & \multicolumn{5}{c}{Single epoch magnitudes} }
\startdata
 & V  1 & V  2 & V  3 & V  4 & V  5 \nl
\tableline
F555W$_{0.5}$  1 & $ 26.00 \pm 0.16 $ & $ 25.77 \pm 0.13 $ & $ 27.20 \pm 0.24 $ & $ 25.36 \pm 0.12 $ & $ 25.93 \pm 0.14 $ \nl
 2 & $ 24.91 \pm 0.12 $ & $ 26.44 \pm 0.15 $ & $ 27.08 \pm 0.23 $ & $ 26.20 \pm 0.15 $ & $ 26.60 \pm 0.20 $ \nl
 3 & $ 25.43 \pm 0.13 $ & $ 25.90 \pm 0.13 $ & $ 26.48 \pm 0.16 $ & $ 25.57 \pm 0.12 $ & $ 26.08 \pm 0.15 $ \nl
 4 & $ 25.27 \pm 0.13 $ & $ 26.15 \pm 0.14 $ & $ 26.52 \pm 0.15 $ & $ 25.71 \pm 0.14 $ & $ 25.25 \pm 0.13 $ \nl
 5 & $ 25.56 \pm 0.16 $ & $ 26.55 \pm 0.18 $ & $ 27.37 \pm 0.32 $ & $ 26.05 \pm 0.16 $ & $ 25.62 \pm 0.13 $ \nl
 6 & $ 25.77 \pm 0.14 $ & $ 26.26 \pm 0.14 $ & $ 27.32 \pm 0.26 $ & $ 26.29 \pm 0.17 $ & $ 25.79 \pm 0.14 $ \nl
 7 & $ 26.19 \pm 0.25 $ & $ 25.72 \pm 0.14 $ & $ 26.67 \pm 0.21 $ & $ 26.61 \pm 0.24 $ & $ 26.37 \pm 0.23 $ \nl
 8 & $ 26.29 \pm 0.14 $ & $ 25.51 \pm 0.16 $ & $ 26.71 \pm 0.21 $ & $ 26.86 \pm 0.20 $ & $ 26.33 \pm 0.17 $ \nl
 9 & $ 25.10 \pm 0.12 $ & $ 26.61 \pm 0.18 $ & $ 26.44 \pm 0.20 $ & $ 26.20 \pm 0.14 $ & $ 26.25 \pm 0.15 $ \nl
10 & $ 25.05 \pm 0.11 $ & $ 26.18 \pm 0.17 $ & $ 26.69 \pm 0.19 $ & $ 26.16 \pm 0.15 $ & $ 26.41 \pm 0.16 $ \nl
11 & $ 25.45 \pm 0.12 $ & $ 25.31 \pm 0.13 $ & $ 27.32 \pm 0.28 $ & $ 26.58 \pm 0.18 $ & $ 25.59 \pm 0.13 $ \nl
12 & $ 25.44 \pm 0.17 $ & $ 25.94 \pm 0.15 $ & $ 27.39 \pm 0.29 $ & $ 25.90 \pm 0.13 $ & $ 25.46 \pm 0.12 $ \nl
13 & $ 26.14 \pm 0.17 $ & $ 25.54 \pm 0.14 $ & $ 27.50 \pm 0.35 $ & $ 26.77 \pm 0.19 $ & $ 26.24 \pm 0.16 $ \nl
14 & $ 25.49 \pm 0.15 $ & $ 25.11 \pm 0.12 $ & $ 26.74 \pm 0.23 $ & $ 26.81 \pm 0.18 $ & $ 25.66 \pm 0.14 $ \nl
15 & $ 26.28 \pm 0.19 $ & $ 25.50 \pm 0.13 $ & $ 26.36 \pm 0.16 $ & $ 26.54 \pm 0.19 $ &      \nodata       \nl
 \nl
F814W$_{0.5}$  1 & $ 24.92 \pm 0.16 $ & $ 25.17 \pm 0.16 $ & $ 26.33 \pm 0.28 $ & $ 24.85 \pm 0.13 $ & $ 24.67 \pm 0.13 $ \nl
 2 & $ 24.43 \pm 0.15 $ & $ 25.66 \pm 0.18 $ & $ 25.75 \pm 0.20 $ & $ 25.34 \pm 0.18 $ & $ 24.97 \pm 0.14 $ \nl
 9 & $ 24.08 \pm 0.13 $ & $ 25.26 \pm 0.17 $ & $ 26.20 \pm 0.34 $ & $ 24.71 \pm 0.13 $ & $ 25.03 \pm 0.19 $ \nl
12 & $ 24.73 \pm 0.21 $ & $ 24.84 \pm 0.14 $ & $ 26.51 \pm 0.41 $ & $ 24.80 \pm 0.13 $ & $ 24.39 \pm 0.12 $ \nl
\tablebreak
 & V  6 & V  7 & V  8 & V  9 & V 10 \nl
\tableline
F555W$_{0.5}$   1 & $ 25.95 \pm 0.16 $ & $ 25.68 \pm 0.09 $ & $ 27.93 \pm 0.46 $ & $ 25.93 \pm 0.10 $ & $ 26.75 \pm 0.21 $ \nl
 2 & $ 26.57 \pm 0.20 $ & $ 25.84 \pm 0.11 $ & $ 27.21 \pm 0.22 $ & $ 26.52 \pm 0.19 $ & $ 26.59 \pm 0.21 $ \nl
 3 & $ 25.35 \pm 0.14 $ & $ 26.09 \pm 0.12 $ & $ 28.21 \pm 0.37 $ & $ 26.60 \pm 0.17 $ & $ 26.11 \pm 0.17 $ \nl
 4 & $ 25.49 \pm 0.15 $ & $ 26.16 \pm 0.13 $ & $ 28.18 \pm 0.50 $ & $ 26.03 \pm 0.10 $ & $ 26.67 \pm 0.20 $ \nl
 5 & $ 25.76 \pm 0.16 $ & $ 26.15 \pm 0.13 $ & $ 27.24 \pm 0.26 $ & $ 26.03 \pm 0.11 $ & $ 27.22 \pm 0.26 $ \nl
 6 & $ 25.68 \pm 0.13 $ & $ 26.12 \pm 0.13 $ & $ 26.89 \pm 0.18 $ & $ 26.26 \pm 0.12 $ & $ 27.15 \pm 0.28 $ \nl
 7 & $ 26.65 \pm 0.26 $ & $ 25.61 \pm 0.10 $ & $ 28.60 \pm 1.13 $ & $ 26.49 \pm 0.19 $ & $ 26.74 \pm 0.23 $ \nl
 8 & $ 26.66 \pm 0.20 $ & $ 25.58 \pm 0.08 $ & $ 27.33 \pm 0.24 $ & $ 27.05 \pm 0.29 $ & $ 27.95 \pm 0.48 $ \nl
 9 & $ 25.31 \pm 0.15 $ & $ 25.91 \pm 0.12 $ & $ 26.93 \pm 0.21 $ & $ 26.13 \pm 0.11 $ & $ 27.59 \pm 0.40 $ \nl
10 & $ 25.43 \pm 0.15 $ & $ 25.89 \pm 0.11 $ & $ 26.48 \pm 0.14 $ & $ 26.19 \pm 0.14 $ & $ 27.56 \pm 0.44 $ \nl
11 & $ 25.58 \pm 0.15 $ & $ 26.40 \pm 0.17 $ & $ 28.41 \pm 0.66 $ & $ 26.58 \pm 0.16 $ & $ 27.46 \pm 0.35 $ \nl
12 & $ 26.32 \pm 0.20 $ & $ 25.71 \pm 0.10 $ & $ 26.81 \pm 0.20 $ & $ 25.91 \pm 0.11 $ & $ 27.43 \pm 0.41 $ \nl
13 & $ 25.94 \pm 0.18 $ & $ 26.39 \pm 0.15 $ & $ 28.28 \pm 0.57 $ & $ 26.55 \pm 0.18 $ & $ 26.74 \pm 0.23 $ \nl
14 &      \nodata       & $ 25.51 \pm 0.08 $ & $ 27.30 \pm 0.23 $ & $ 26.58 \pm 0.17 $ & $ 27.43 \pm 0.37 $ \nl
15 & $ 26.53 \pm 0.18 $ & $ 26.06 \pm 0.12 $ & $ 27.69 \pm 0.41 $ & $ 26.45 \pm 0.17 $ & $ 28.23 \pm 0.62 $ \nl
 \nl
F814W$_{0.5}$  1 & $ 25.29 \pm 0.21 $ & $ 24.49 \pm 0.08 $ & $ 26.70 \pm 0.41 $ & $ 24.71 \pm 0.10 $ & $ 25.70 \pm 0.23 $ \nl
 2 & $ 25.15 \pm 0.17 $ & $ 24.77 \pm 0.09 $ & $ 26.06 \pm 0.22 $ & $ 25.07 \pm 0.12 $ & $ 25.45 \pm 0.23 $ \nl
 9 & $ 25.50 \pm 0.26 $ & $ 24.56 \pm 0.14 $ & $ 26.37 \pm 0.42 $ & $ 24.88 \pm 0.12 $ & $ 25.98 \pm 0.36 $ \nl
12 & $ 25.29 \pm 0.21 $ & $ 24.56 \pm 0.09 $ & $ 25.68 \pm 0.18 $ & $ 24.62 \pm 0.09 $ & $ 26.80 \pm 0.63 $ \nl
\tablebreak
 & V 11 & V 12 & V 13 & ~~~ & ~~~ \nl
\tableline
F555W$_{0.5}$ 1 & $ 25.71 \pm 0.15 $ & $ 26.29 \pm 0.20 $ & $ 25.93 \pm 0.16 $ & ~~~ ~~~~ ~~~  & ~~~ ~~~~ ~~~ \nl
 2 & $ 26.40 \pm 0.18 $ & $ 25.39 \pm 0.14 $ & $ 26.66 \pm 0.26 $ \nl
 3 & $ 26.02 \pm 0.18 $ & $ 25.91 \pm 0.17 $ & $ 25.36 \pm 0.13 $ \nl
 4 & $ 26.01 \pm 0.17 $ & $ 25.94 \pm 0.18 $ & $ 25.40 \pm 0.13 $ \nl
 5 & $ 26.25 \pm 0.18 $ & $ 26.08 \pm 0.20 $ & $ 25.53 \pm 0.13 $ \nl
 6 & $ 26.42 \pm 0.29 $ & $ 25.56 \pm 0.16 $ & $ 25.96 \pm 0.16 $ \nl
 7 & $ 26.40 \pm 0.23 $ & $ 25.39 \pm 0.15 $ & $ 25.90 \pm 0.21 $ \nl
 8 & $ 26.05 \pm 0.15 $ & $ 25.51 \pm 0.14 $ & $ 26.10 \pm 0.17 $ \nl
 9 & $ 26.87 \pm 0.26 $ & $ 26.15 \pm 0.21 $ & $ 26.41 \pm 0.21 $ \nl
10 & $ 26.55 \pm 0.22 $ & $ 26.09 \pm 0.22 $ & $ 26.28 \pm 0.22 $ \nl
11 & $ 25.79 \pm 0.14 $ & $ 25.31 \pm 0.13 $ & $ 25.60 \pm 0.13 $ \nl
12 & $ 26.29 \pm 0.17 $ & $ 25.79 \pm 0.17 $ & $ 25.86 \pm 0.14 $ \nl
13 & $ 25.89 \pm 0.15 $ & $ 25.55 \pm 0.16 $ & $ 26.08 \pm 0.20 $ \nl
14 & $ 25.81 \pm 0.14 $ & $ 26.12 \pm 0.22 $ & $ 25.51 \pm 0.14 $ \nl
15 & $ 26.20 \pm 0.23 $ & $ 25.86 \pm 0.25 $ & $ 26.60 \pm 0.21 $ \nl
 \nl
F814W$_{0.5}$  1 & $ 24.96 \pm 0.17 $ & $ 25.10 \pm 0.18 $ & $ 24.98 \pm 0.20 $ \nl
 2 & $ 25.48 \pm 0.23 $ & $ 24.41 \pm 0.14 $ & $ 25.06 \pm 0.17 $ \nl
 9 & $ 25.15 \pm 0.21 $ & $ 24.94 \pm 0.20 $ & $ 25.08 \pm 0.23 $ \nl
12 & $ 25.18 \pm 0.20 $ & $ 24.75 \pm 0.17 $ & $ 24.68 \pm 0.17 $ \nl
 \nl
\enddata
\end{deluxetable}

\clearpage

%Produced by n7331_ceph_plot_pap.f
\begin{deluxetable}{rccccc}
\tablecaption{Calibrated ALLFRAME F555W$_{0.5}$ and F814W$_{0.5}$ magnitudes 
of Cepheids with good light curves. 
\label{good_cephotH}}
\tablewidth{16cm}
\tableheadfrac{0.0}
\tablehead{Epoch & \multicolumn{5}{c}{Single epoch magnitudes} }
\startdata
 & V  1 & V  2 & V  3 & V  4 & V  5 \nl
\tableline
F555W$_{0.5}$   1 & $ 26.09 \pm 0.26 $ & $ 25.66 \pm 0.13 $ & $ 27.30 \pm 0.28 $ & $ 25.65 \pm 0.19 $ & $ 25.97 \pm 0.14 $ \nl
 2 & $ 24.97 \pm 0.11 $ & $ 26.11 \pm 0.18 $ & $ 26.70 \pm 0.23 $ & $ 26.19 \pm 0.13 $ & $ 26.39 \pm 0.17 $ \nl
 3 & $ 25.42 \pm 0.12 $ &      \nodata       & $ 26.63 \pm 0.20 $ & $ 25.52 \pm 0.13 $ & $ 26.20 \pm 0.21 $ \nl
 4 &      \nodata       & $ 25.81 \pm 0.16 $ & $ 26.63 \pm 0.23 $ & $ 25.66 \pm 0.13 $ & $ 25.52 \pm 0.17 $ \nl
 5 &      \nodata       & $ 26.41 \pm 0.25 $ & $ 27.23 \pm 0.35 $ & $ 26.05 \pm 0.16 $ & $ 25.76 \pm 0.11 $ \nl
 6 & $ 25.49 \pm 0.90 $ & $ 26.13 \pm 0.21 $ & $ 27.42 \pm 0.30 $ & $ 26.28 \pm 0.16 $ & $ 26.04 \pm 0.19 $ \nl
 7 & $ 25.90 \pm 0.33 $ & $ 25.62 \pm 0.19 $ & $ 26.52 \pm 0.20 $ & $ 26.39 \pm 0.31 $ & $ 26.17 \pm 0.25 $ \nl
 8 & $ 26.05 \pm 0.18 $ & $ 25.38 \pm 0.16 $ & $ 26.78 \pm 0.21 $ & $ 27.00 \pm 0.25 $ & $ 26.25 \pm 0.17 $ \nl
 9 & $ 25.01 \pm 0.10 $ & $ 26.21 \pm 0.19 $ &      \nodata       & $ 26.16 \pm 0.21 $ & $ 26.32 \pm 0.14 $ \nl
10 & $ 25.05 \pm 0.13 $ & $ 25.96 \pm 0.47 $ & $ 26.66 \pm 0.22 $ & $ 26.22 \pm 0.17 $ & $ 26.52 \pm 0.19 $ \nl
11 & $ 25.29 \pm 0.13 $ & $ 25.15 \pm 0.20 $ & $ 27.32 \pm 0.27 $ & $ 26.54 \pm 0.18 $ & $ 25.70 \pm 0.13 $ \nl
12 & $ 25.31 \pm 0.13 $ & $ 25.91 \pm 0.11 $ & $ 27.86 \pm 0.43 $ & $ 25.99 \pm 0.15 $ & $ 25.57 \pm 0.13 $ \nl
13 & $ 26.17 \pm 0.15 $ & $ 25.47 \pm 0.12 $ & $ 27.67 \pm 0.49 $ & $ 27.03 \pm 0.22 $ & $ 26.48 \pm 0.16 $ \nl
14 & $ 25.05 \pm 0.12 $ & $ 25.22 \pm 0.13 $ & $ 26.76 \pm 0.20 $ & $ 26.85 \pm 0.32 $ & $ 25.67 \pm 0.21 $ \nl
15 & $ 26.16 \pm 0.18 $ & $ 25.69 \pm 0.16 $ & $ 26.48 \pm 0.15 $ & $ 26.49 \pm 0.26 $ & $ 27.01 \pm 0.29 $ \nl
 \nl
F814W$_{0.5}$  1 & $ 24.86 \pm 0.14 $ & $ 25.08 \pm 0.14 $ & $ 26.22 \pm 0.23 $ & $ 24.86 \pm 0.12 $ & $ 24.67 \pm 0.07 $ \nl
 2 & $ 24.20 \pm 0.09 $ & $ 25.45 \pm 0.16 $ & $ 25.71 \pm 0.19 $ & $ 25.20 \pm 0.10 $ & $ 25.02 \pm 0.11 $ \nl
 8 & $ 24.06 \pm 0.13 $ & $ 25.16 \pm 0.24 $ & $ 26.33 \pm 0.39 $ & $ 24.78 \pm 0.13 $ & $ 25.39 \pm 0.23 $ \nl
12 & $ 24.33 \pm 0.15 $ & $ 25.08 \pm 0.12 $ & $ 26.67 \pm 0.36 $ & $ 25.04 \pm 0.13 $ & $ 24.81 \pm 0.11 $ \nl
\tablebreak
 & V  6 & V  7 & V  8 & V  9 & V 10 \nl
\tableline
F555W$_{0.5}$  & $ 25.55 \pm 0.14 $ & $ 25.62 \pm 0.12 $ & $ 28.02 \pm 0.58 $ & $ 25.89 \pm 0.13 $ & $ 27.00 \pm 0.31 $ \nl
 2 & $ 25.97 \pm 0.24 $ & $ 26.00 \pm 0.19 $ & $ 27.40 \pm 0.37 $ & $ 26.57 \pm 0.38 $ & $ 26.79 \pm 0.32 $ \nl
 3 & $ 25.05 \pm 0.11 $ & $ 26.26 \pm 0.18 $ & $ 27.53 \pm 0.44 $ & $ 25.89 \pm 0.22 $ & $ 26.60 \pm 0.34 $ \nl
 4 & $ 25.07 \pm 0.12 $ & $ 26.16 \pm 0.16 $ & $ 28.38 \pm 0.74 $ & $ 26.05 \pm 0.12 $ & $ 26.75 \pm 0.27 $ \nl
 5 & $ 25.42 \pm 0.13 $ & $ 26.15 \pm 0.13 $ & $ 27.43 \pm 0.37 $ & $ 26.06 \pm 0.16 $ & $ 27.36 \pm 0.45 $ \nl
 6 & $ 25.54 \pm 0.14 $ & $ 26.10 \pm 0.15 $ & $ 27.01 \pm 0.24 $ & $ 26.38 \pm 0.18 $ & $ 27.27 \pm 0.34 $ \nl
 7 & $ 26.03 \pm 0.24 $ & $ 25.56 \pm 0.19 $ &      \nodata       & $ 26.44 \pm 0.26 $ & $ 26.70 \pm 0.26 $ \nl
 8 & $ 25.96 \pm 0.33 $ & $ 25.71 \pm 0.11 $ & $ 27.40 \pm 0.34 $ & $ 27.03 \pm 0.35 $ & $ 27.61 \pm 0.40 $ \nl
 9 & $ 24.92 \pm 0.11 $ & $ 25.88 \pm 0.17 $ & $ 26.51 \pm 1.13 $ & $ 25.98 \pm 0.15 $ & $ 27.68 \pm 0.51 $ \nl
10 & $ 25.02 \pm 0.11 $ & $ 26.00 \pm 0.10 $ & $ 26.64 \pm 0.22 $ & $ 26.25 \pm 0.22 $ & $ 27.68 \pm 0.51 $ \nl
11 & $ 25.40 \pm 0.14 $ & $ 26.28 \pm 0.24 $ & $ 28.18 \pm 0.59 $ & $ 26.52 \pm 0.19 $ & $ 27.21 \pm 0.34 $ \nl
12 & $ 26.36 \pm 0.23 $ & $ 25.70 \pm 0.14 $ & $ 26.92 \pm 0.23 $ & $ 25.82 \pm 0.14 $ & $ 27.53 \pm 0.36 $ \nl
13 & $ 25.61 \pm 0.16 $ & $ 26.43 \pm 0.22 $ & $ 27.83 \pm 0.77 $ & $ 26.45 \pm 0.19 $ & $ 26.90 \pm 0.22 $ \nl
14 & $ 24.97 \pm 0.11 $ & $ 25.61 \pm 0.13 $ & $ 27.60 \pm 0.30 $ & $ 26.48 \pm 0.23 $ & $ 27.43 \pm 0.38 $ \nl
15 & $ 26.45 \pm 0.21 $ & $ 26.12 \pm 0.17 $ & $ 27.77 \pm 0.50 $ & $ 26.60 \pm 0.26 $ & $ 28.26 \pm 0.86 $ \nl
 \nl
F814W$_{0.5}$  1 & $ 24.58 \pm 0.10 $ & $ 24.65 \pm 0.16 $ & $ 26.83 \pm 0.44 $ & $ 24.84 \pm 0.16 $ & $ 25.75 \pm 0.19 $ \nl
 2 & $ 24.94 \pm 0.13 $ & $ 24.87 \pm 0.16 $ & $ 26.20 \pm 0.34 $ & $ 25.10 \pm 0.13 $ & $ 25.66 \pm 0.22 $ \nl
 8 & $ 25.19 \pm 0.22 $ & $ 24.74 \pm 0.17 $ & $ 26.76 \pm 0.44 $ & $ 25.06 \pm 0.21 $ & $ 26.09 \pm 0.33 $ \nl
12 & $ 25.16 \pm 0.13 $ & $ 24.76 \pm 0.15 $ & $ 25.97 \pm 0.25 $ & $ 24.73 \pm 0.12 $ & $ 26.91 \pm 0.72 $ \nl
\tablebreak
 & V 11 & V 12 & V 13 \nl
\tableline
F555W$_{0.5}$ 1 & $ 25.95 \pm 0.20 $ & $ 26.29 \pm 0.25 $ & $ 26.23 \pm 0.32 $ & ~~~ ~~~~ ~~~ & ~~~ ~~~~ ~~~  \nl
 2 & $ 26.65 \pm 0.22 $ & $ 25.32 \pm 0.10 $ & $ 26.78 \pm 0.25 $ \nl
 3 & $ 26.20 \pm 0.33 $ & $ 25.96 \pm 0.18 $ & $ 25.39 \pm 0.14 $ \nl
 4 & $ 26.10 \pm 0.16 $ & $ 25.94 \pm 0.16 $ & $ 25.44 \pm 0.20 $ \nl
 5 & $ 26.26 \pm 0.24 $ & $ 25.95 \pm 0.17 $ & $ 25.58 \pm 0.13 $ \nl
 6 & $ 26.64 \pm 0.24 $ & $ 25.53 \pm 0.15 $ & $ 25.83 \pm 0.14 $ \nl
 7 & $ 26.69 \pm 0.38 $ & $ 25.36 \pm 0.16 $ & $ 25.99 \pm 0.33 $ \nl
 8 & $ 25.88 \pm 0.19 $ & $ 25.56 \pm 0.23 $ & $ 26.22 \pm 0.17 $ \nl
 9 & $ 26.97 \pm 0.41 $ & $ 26.03 \pm 0.22 $ & $ 26.51 \pm 0.17 $ \nl
10 & $ 26.80 \pm 0.25 $ & $ 26.11 \pm 0.26 $ & $ 26.47 \pm 0.17 $ \nl
11 & $ 25.81 \pm 0.15 $ & $ 25.16 \pm 0.15 $ & $ 25.61 \pm 0.16 $ \nl
12 & $ 26.50 \pm 0.23 $ & $ 26.07 \pm 0.30 $ & $ 26.08 \pm 0.19 $ \nl
13 & $ 26.24 \pm 0.20 $ & $ 25.53 \pm 0.17 $ & $ 26.42 \pm 0.25 $ \nl
14 & $ 25.90 \pm 0.13 $ & $ 26.26 \pm 0.19 $ & $ 25.67 \pm 0.15 $ \nl
15 & $ 26.46 \pm 0.23 $ & $ 26.23 \pm 0.22 $ & $ 26.58 \pm 0.31 $ \nl
 \nl
F814W$_{0.5}$  1 & $ 25.17 \pm 0.15 $ & $ 25.36 \pm 0.17 $ & $ 25.20 \pm 0.17 $ \nl
 2 & $ 25.67 \pm 0.29 $ & $ 24.59 \pm 0.10 $ & $ 25.20 \pm 0.14 $ \nl
 8 & $ 25.39 \pm 0.22 $ & $ 25.02 \pm 0.20 $ & $ 25.01 \pm 0.24 $ \nl
12 & $ 25.55 \pm 0.24 $ & $ 24.98 \pm 0.14 $ & $ 24.91 \pm 0.15 $ \nl
 \nl
\enddata
\end{deluxetable}

\clearpage

\begin{landscape}

%Produced by n7331_ceph_plot_pap.f
\begin{deluxetable}{lcrrrrrrrrrr}
\tablecaption{Parameters for Cepheids Detected in NGC~7331. \label{good_ceph_tab}}
%\small
\tablewidth{0pt}
\tablehead{
\colhead{ID} & \colhead{Chip} & \multicolumn{2}{c}{RA ~~  (J2000) ~~ Dec}  & 
\colhead{X}   & \colhead{Y} & \colhead{P$_{Alf}$} & 
\colhead{$\langle V \rangle_{Alf}$} & \colhead{$\langle V-I \rangle_{Alf}$} & 
\colhead{P$_{Doph}$} & 
\colhead{$\langle V \rangle_{Doph}$} & \colhead{$\langle V-I \rangle_{Doph}$} }
\startdata
V  1 & 2 & 22:36:59.61 & +34:28:20.1 &  296.3 &   75.1 & 42.59 & $ 25.40 \pm 0.06 $ & $  1.08 \pm 0.09 $ & 42.18 & $ 25.53 \pm 0.04 $ & $  0.89 \pm 0.09 $ \nl
V  2 & 2 & 22:36:56.52 & +34:28:42.2 &  661.8 &  322.1 & 21.19 & $ 25.68 \pm 0.06 $ & $  0.50 \pm 0.10 $ & 21.35 & $ 25.81 \pm 0.04 $ & $  0.54 \pm 0.09 $ \nl
V  3 & 2 & 22:36:58.30 & +34:28:45.9 &  437.8 &  340.7 & 13.91 & $ 26.93 \pm 0.08 $ & $  0.95 \pm 0.16 $ & 13.91 & $ 26.91 \pm 0.06 $ & $  0.94 \pm 0.17 $ \nl
V  4 & 2 & 22:36:58.82 & +34:28:47.4 &  372.4 &  350.4 & 22.60 & $ 26.13 \pm 0.05 $ & $  1.20 \pm 0.08 $ & 22.60 & $ 26.14 \pm 0.04 $ & $  1.12 \pm 0.08 $ \nl
V  5 & 2 & 22:37:00.23 & +34:29:26.7 &  162.1 &  732.2 & 33.94 & $ 26.16 \pm 0.05 $ & $  1.10 \pm 0.08 $ & 33.87 & $ 26.02 \pm 0.04 $ & $  1.30 \pm 0.09 $ \nl
V  6 & 2 & 22:36:59.92 & +34:29:28.2 &  199.8 &  750.4 & 29.05 & $ 25.62 \pm 0.05 $ & $  0.91 \pm 0.08 $ & 29.43 & $ 26.02 \pm 0.06 $ & $  0.81 \pm 0.12 $ \nl
V  7 & 3 & 22:37:01.81 & +34:28:41.3 &  265.1 &   88.9 & 39.90 & $ 25.89 \pm 0.04 $ & $  1.08 \pm 0.09 $ & 39.72 & $ 25.86 \pm 0.03 $ & $  1.22 \pm 0.06 $ \nl
V  8 & 3 & 22:37:02.74 & +34:28:51.7 &  360.3 &  212.1 & 11.13 & $ 27.37 \pm 0.13 $ & $  0.97 \pm 0.23 $ & 11.16 & $ 27.47 \pm 0.12 $ & $  1.40 \pm 0.20 $ \nl
V  9 & 3 & 22:37:03.13 & +34:28:57.1 &  410.6 &  265.2 & 24.41 & $ 26.20 \pm 0.06 $ & $  1.27 \pm 0.10 $ & 25.83 & $ 26.37 \pm 0.04 $ & $  1.52 \pm 0.07 $ \nl
V 10 & 4 & 22:37:03.41 & +34:28:07.9 &  261.1 &  163.3 & 11.61 & $ 27.02 \pm 0.11 $ & $  1.05 \pm 0.25 $ & 11.61 & $ 26.96 \pm 0.09 $ & $  1.04 \pm 0.22 $ \nl
V 11 & 4 & 22:37:04.53 & +34:27:27.8 &  371.8 &  575.5 & 19.82 & $ 26.28 \pm 0.06 $ & $  0.93 \pm 0.14 $ & 19.86 & $ 26.15 \pm 0.05 $ & $  0.98 \pm 0.11 $ \nl
V 12 & 4 & 22:37:04.53 & +34:27:25.6 &  369.7 &  597.4 & 24.82 & $ 25.68 \pm 0.05 $ & $  0.86 \pm 0.09 $ & 26.68 & $ 25.65 \pm 0.05 $ & $  0.99 \pm 0.10 $ \nl
V 13 & 4 & 22:37:03.80 & +34:27:06.9 &  266.6 &  779.3 & 41.20 & $ 25.92 \pm 0.05 $ & $  1.27 \pm 0.10 $ & 37.54 & $ 25.92 \pm 0.05 $ & $  1.16 \pm 0.11 $ \nl
\nl
\enddata
\end{deluxetable}

\end{landscape}

\clearpage

%Produced by pl_plot_lmc_n7331.f
\begin{deluxetable}{lrcccc}
\tablecaption{Fits to PL relations. \label{PL_fit_tab}}
\tablewidth{0pt}
\tablehead{
\colhead{Sample} & \colhead{N}   & \colhead{DM($V$)} & \colhead{DM($I$)} &
\colhead{$E(V-I)^a$)} & \colhead{DM$_0^a$} }
\startdata
\cutinhead{ALLFRAME}  \nl
{\bf All} & 13 & $ 31.36 \pm 0.07 $ & $ 31.17 \pm 0.04 $ & $ 0.19 \pm 0.06$ & $ 30.89 \pm 0.08 $ \nl
P $>$ 15  & 10 & $ 31.32 \pm 0.07 $ & $ 31.15 \pm 0.04 $ & $ 0.17 \pm 0.07$ & $ 30.90 \pm 0.08 $ \nl
P $>$ 20  &  9 & $ 31.33 \pm 0.07 $ & $ 31.15 \pm 0.04 $ & $ 0.18 \pm 0.08$ & $ 30.90 \pm 0.08 $ \nl
P $>$ 25  &  5 & $ 31.52 \pm 0.08 $ & $ 31.31 \pm 0.05 $ & $ 0.21 \pm 0.05$ & $ 31.01 \pm 0.09 $ \nl
\nl
\cutinhead{DoPHOT}  \nl
{\bf All} & 13 & $ 31.41 \pm 0.08 $ & $ 31.16 \pm 0.06 $ & $ 0.25 \pm 0.07$ & $ 30.80 \pm 0.14 $ \nl
P $>$ 15  & 10 & $ 31.38 \pm 0.10 $ & $ 31.18 \pm 0.07 $ & $ 0.21 \pm 0.09$ & $ 30.88 \pm 0.16 $ \nl
P $>$ 20  &  9 & $ 31.41 \pm 0.10 $ & $ 31.20 \pm 0.08 $ & $ 0.21 \pm 0.10$ & $ 30.90 \pm 0.18 $ \nl
P $>$ 25  &  7 & $ 31.50 \pm 0.10 $ & $ 31.24 \pm 0.09 $ & $ 0.26 \pm 0.10$ & $ 30.86 \pm 0.20 $ \nl
\nl
\enddata
\tablecomments{
(a) Due to the finite width of the Cepheid instability strip, errors in DM($V$) and DM($I$) are partially
correlated when calculating the errors in $E(V-I)$ and DM$_0$.}
\end{deluxetable}

\clearpage

\begin{deluxetable}{lrl}
\tablecaption{Error Budget. \label{Error_tab}}
\tablewidth{0pt}
\tablehead{
\colhead{Source of Uncertainty} & \colhead{Error (mag)} & 
\colhead{Notes} }
\startdata
%{\bf LMC} & {\bf CEPHEID PL CALIBRATION} &  \\
{\bf CEPHEID PL CALIBRATION} &  &  \nl
(a) LMC True Moduus 		& $\pm$ 0.10 	& (1) \nl
(b) V PL Zero Point 		& $\pm$ 0.05 	& (2),(3) \nl
(c) I PL Zero Point 		& $\pm$ 0.03 	& (2),(4) \nl
(S1) Systematic Uncertainty 	& $\pm$ 0.12 
   	& (a),(b),(c) combined in quadrature \nl
\nl
{\bf NGC 7331 MODULUS} 		& 		&  \nl
(d) HST V-Band Zero Point 	& $\pm$ 0.04 	& (5) \nl
(e) HST I-Band Zero Point 	& $\pm$ 0.04 	& (5) \nl
(R1) Cepheid True Modulus 	& $\pm$ 0.11 	& (6)  \nl
(f) PL Fit (V) (ALLFRAME)	& $\pm$ 0.07 	& (7) \nl
(g) PL Fit (I) (ALLFRAME)	& $\pm$ 0.04 	& (7) \nl
(R2) Cepheid True Modulus 	& $\pm$ 0.08    & (f),(g) partially correlated,
(8) \nl
(S2) Metallicity Uncertainty 	& +0.05 $\pm$ 0.04 
  	& See text for details \nl
\nl
{\bf TOTAL UNCERTAINTY} 	& 		& \nl
(R) Random Errors 		& $\pm$ 0.14 	
	& (R1) and (R2) combined in quadrature \nl
(S) Systematic Errors 		& +0.05 $\pm$ 0.13 
	& (S1) and (S2) combined in quadrature \nl
\nl
\enddata
\tablecomments{
(1) Adopted from Westerlund (1997). \newline
(2) Derived from the observed scatter in the Madore \& Freedman (1991) 
PL relation, with 32  contributing LMC Cepheids. \newline
(3) V-band scatter: $\pm$ 0.27. \newline
(4) I-band scatter: $\pm$ 0.18. \newline
(5) Contributing uncertainties from aperture corrections,
the Holtzman et al. (1995) zero points, and the long-versus short uncertainty,
combined in quadrature.  
Adopted aperture correction contribution is the worst case formal 
uncertainty ($\pm$0.02 mag) for the NGC~7331 aperture corrections.
Adopted Holtzman et al. (1995) zero point uncertainty is $\pm$0.02 mag.
Adopted long-versus-short-exposure correction uncertainty is $\pm$0.02 mag. \newline
(6) Assuming that photometric errors (d,e) are
uncorrelated between filters, and noting that the V and I magnitudes are
multiplied by +1.45 and -2.45, respectively, when correcting for reddening, 
results in a derived error on the true modulus of 
[(1.45)$^{2}$(0.04)$^{2}$+(-2.45)$^{2}$(0.04)$^{2}$]$^{1/2}$ $=$ 0.11 mag. \newline
(7)  Uncertainties for the mean apparent V and I moduli are limited by the 
apparent width of the derived PL relation, reduced by the population size of 
contributing Cepheids for NGC~7331 (13 variables).  
Contributing effects include photometric errors,  differential
reddening and intrinsic strip filling.  
(8)  The partially correlated nature of the derived PL width uncertainties 
is taken into account by the (correlated) dereddening procedure, coupled 
with the largely ``degenerate-with-reddening'' positioning of individual 
Cepheids within the instability strip.
}
\end{deluxetable}

\clearpage

\begin{deluxetable}{lcl}
\tablecaption{Comparison of Published Distances to NGC~7331. \label{Dist_tab}}
\tablewidth{0pt}
\tablehead{
\colhead{Method} & \colhead{Distance (Mpc)}   & \colhead{Reference} }
\startdata
Sizes of HII regions	  & ~9.1 & Sersic 1960 \nl
Luminosity Class	  & ~7.9 & van den Bergh 1960 \nl
H$_0 = 75$  		  & 14.4 & Rubin et al.\ 1965 \nl
M31/NGC~7331 diameters    & 15.4 & Rubin et al.\ 1965 \nl
Mass estimators		  & 13.0 & Balkowski et al.\ 1973 \nl
Tully-Fisher ($H$\ band)  & 10.1 & Aaronson et al.\ 1980, 1982a \nl
Velocity ratio with Virgo & 12.0 & Aaronson et al.\ 1982a \nl
Sizes of dark clouds	  & ~7.2 & Osman et al.\ 1982 \nl
Tully-Fisher ($B$\ band)  & ~9.6 & Bottinelli et al.\ 1985 \nl
H$_0 = 75$ 		  & 14.3 & Tully 1988 \nl
Cepheids 		  & 15.1 & This paper (ALLFRAME) \nl
\nl
\enddata
\end{deluxetable}

\clearpage

\begin{deluxetable}{lcccl}
\tablecaption{IRTF Calibration. \label{TF_tab}}
\tablewidth{0pt}
\tablehead{\colhead{Galaxy} & \colhead{DM} & 
\colhead{log$\Delta$V} & \colhead{$H^{abs}_{g}$} &
\colhead{Source of Distance} }
\startdata
 NGC 224 (M31)  & 24.40 & 2.751 & -23.93 & Freedman \& Madore 1990  \nl
 NGC 300        & 26.66 & 2.371 & -19.99 & Madore \& Freedman 1991  \nl
 NGC 598 (M33)  & 24.50 & 2.392 & -20.50 & Freedman 1990  \nl
 NGC 925        & 29.84 & 2.422 & -21.48 & Silbermann et al. 1996 \nl
 NGC 2090       & 30.45 & 2.529 & -21.98 & Phelps et al. 1998 \nl
 NGC 2403       & 27.50 & 2.488 & -21.29 & Freedman 1990 \nl
 NGC 3031 (M81) & 27.80 & 2.716 & -23.62 & Freedman et al. 1994 \nl
 NGC 3351 (M95) & 30.01 & 2.573 & -22.72 & Graham et al. 1997 \nl
 NGC 3368 (M96) & 30.32 & 2.709 & -23.60 & Tanvir et al. 1995 \nl
 NGC 3621       & 29.13 & 2.536 & -22.43 & Rawson et al. 1997 \nl
 NGC 4536       & 31.10 & 2.576 & -22.99 & Saha et al. 1996 \nl
 NGC 7331       & 30.89 & 2.740 & -24.81 & This paper  \nl
\nl
\enddata
\tablecomments{$\Delta$V are the 21 cm line widths in km/s, as listed in 
Table 4 of Tormen \& Burstein (1995), which are from Aaronson et al 
(1982a), and corrected for inclination and z, but without any 3\deg\ 
addition term to the inclinations. 
$H_{g}$ mags are from Tormen \& Burstein (1995), which are the Aaronson 
et al. (1982a) $H_{0.5}$ mags recalculated in accordance with the isophotal 
diameters of RC3, and uncorrected for inclination.}
\end{deluxetable}

\clearpage

\begin{deluxetable}{lrrcc}
\tablecaption{NGC~7331 Group. \label{Group}}
\tablewidth{0pt}
\tablehead{\colhead{Galaxy} & \multicolumn{2}{c}{RA ~~  (J2000) ~~ Dec}  & 
\colhead{V$_{\rm hel}$ km s$^{-1}$} & \colhead{R (degrees)}}
\startdata
NGC 7217   & 22:07:52.1 & +31:21:35 &    1195 &  6.85  \nl 
NGC 7292   & 22:28:25.0 & +30:17:30 &    1226 &  4.50  \nl 
2228+3300  & 22:29:36.8 & +33:17:17 &    1130 &  1.92  \nl 
UGC 12060  & 22:30:33.9 & +33:49:11 &    1128 &  1.47  \nl 
UGC 12082  & 22:34:10.9 & +32:51:31 &    1047 &  1.66  \nl 
NGC 7320A  & 22:36:03.4 & +33:56:54 &    1029 &  0.51  \nl 
NGC 7331   & 22:37:04.0 & +34:24:57 &    1064 &  0.00  \nl 
UGC 12212  & 22:50:29.5 & +29:08:15 &    1120 &  6.00  \nl 
NGC 7457   & 23:00:59.8 & +30:08:39 &    1054 &  6.63  \nl 
UGC 12311  & 23:01:23.8 & +30:14:08 &    1150 &  6.62  \nl 
UGC 12404  & 23:11:04.8 & +29:38:39 &    1066 &  8.64  \nl 
\nl
\enddata
\end{deluxetable}

\clearpage

\begin{deluxetable}{crrrrrr}
\tablenum{A1}
\tablecaption{DoPHOT Aperture Correction Spatial Coefficients. \label{App_AC_doph}}
\tablewidth{0pt}
\tablehead{
\colhead{Chip} & \colhead{C0} & \colhead{C1} & \colhead{C2} 
& \colhead{C3} & \colhead{C4} & \colhead{C5} }
\startdata
\cutinhead{F555W} \nl
1 &  $-$0.99431 & 7.222919E-5 & 3.826290E-4 & $-$7.732423E-8 & $-$4.587308E-7
& $-$7.685230E-9 \nl
2 &     $-$0.817943  & 1.941904E-4 &
 3.665115E-4 &
$-$3.065632E-7 &
$-$4.345170E-7 &
 1.143239E-7 \nl
3 &     $-$0.73354  &2.550644E-5 &
 3.484296E-4 &
$-$7.614571E-8 &
$-$4.447892E-7 &
 1.363507E-7 \nl
4 &     $-$0.818136 & 2.384645E-4 &
 4.255572E-4 &
$-$3.253283E-7 &
$-$6.087260E-7 &
 7.968657E-8 \nl
\nl
\cutinhead{F814W} \nl
1 &      0.550125 & 7.906689E-5 &
 1.449731E-4 &
$-$2.173195E-7 &
$-$2.504804E-7 &
 2.752383E-7 \nl

2 &     $-$0.914758 & 1.835472E-4 &
 2.947996E-4 &
$-$2.878237E-7 &
$-$3.648580E-7 &
 1.434279E-7 \nl

3 &     $-$0.884129 & 7.124484E-6 &
 2.373478E-4 &
$-$1.044799E-7 &
$-$3.485253E-7 &
 2.285843E-7 \nl

4 &     $-$0.949892 & 2.751133E-4 &
 3.490840E-4 &
$-$2.675065E-7 &
$-$3.724136E-7 &
$-$1.041973E-7 \nl
\enddata
\end{deluxetable}

\clearpage

\begin{deluxetable}{ccccc}
\tablenum{A2}
\tablecaption{Zero-points and Aperture Corrections. \label{App_ZP_AC}}
\tablewidth{0pt}
\tablehead{
\colhead{Filter} & \colhead{Chip PC1} & \colhead{Chip WF2} & 
\colhead{Chip WF3} & \colhead{Chip WF4} }
\startdata
\cutinhead{DoPHOT Zero-points} \nl
F555W  &  $-$7.630  &  $-$7.536  &  $-$7.554  &  $-$7.559 \nl
F814W  &  $-$8.536  &  $-$8.381  &  $-$8.436  &  $-$8.441 \nl
\nl
\cutinhead{ALLFRAME Mean Aperture Corrections} \nl
F555W &  $-$0.062 & +0.031 & +0.029 & +0.058 \nl
F814W &  $-$0.051 & +0.069 & +0.036 & +0.070 \nl
\nl
\cutinhead{ALLFRAME Zero-points} \nl
F555W  &  $-$0.967  &  $-$0.958  &  $-$0.950  & $-$0.973 \nl
F814W  &  $-$1.861  &  $-$1.823  &  $-$1.842  & $-$1.870 \nl
\enddata
\end{deluxetable}

\clearpage

%\begin{figure}
%\figurenum{\label{INT_fig}}
%\plotone{n7331_f01.eps}
%\end{figure}

%\begin{figure}
%\figurenum{\label{wfpc2_cephs}}
%\plotone{n7331_f02.eps}
%\end{figure}

\begin{figure}
\figurenum{\label{doC_alf_comp_fig}}
\plotone{n7331_f03.eps}
\end{figure}

\begin{figure}
\figurenum{\label{CMD_fig}}
\plotone{n7331_f04.eps}
\end{figure}

\begin{figure}
\figurenum{\label{LF_fig}}
\plotone{n7331_f05.eps}
\end{figure}

\begin{figure}
\figurenum{\label{FC_fig}}
\plotone{n7331_f06.eps}
\end{figure}

%\begin{figure}
%\figurenum{\label{lc_fig}}
%\plotone{n7331_f07.eps}
%\end{figure}

\begin{figure}
\figurenum{\label{PL_figH}}
\plotone{n7331_f08.eps}
\end{figure}

\begin{figure}
\figurenum{\label{PL_figD}}
\plotone{n7331_f09.eps}
\end{figure}

\begin{figure}
\figurenum{\label{TF_fig}}
\plotone{n7331_f10.eps}
\end{figure}


\newpage

\begin{references}
 \reference{aaro80} Aaronson,~M., Mould,~J., \& Huchra,~J. 1980, \apj, 237, 655
 \reference{aaro82a} Aaronson,~M., et al. 1982a, \apjs, 50, 241
 \reference{aaro82b} Aaronson,~M., et al. 1982b, \apj, 258, 64
 \reference{aaro83} Aaronson,~M., \& Mould,~J.~R. 1983, \apj, 265, 1
 \reference{balk73} Balkowski,~C., et al.\ 1973, \aa, 25, 319
 \reference{bege87} Begeman,~K.~G. 1987, PhD thesis, University of Groningen
%\reference{bege91} Begeman,~K.~G., Broeils,~A.~H., \& Sanders,~R.~H. 1991, 
\mnras, 249, 523
%\reference{bosm81} Bosma,~A. 1981, \aj, 86, 1791
 \reference{bott85} Bottinelli,~L., et al.\ 1985, \aaps, 59, 43
 \reference{bowe93} Bower,~G.~A., Richstone,~D.~O., Bothun,~G.~D., \& 
Heckman,~T.~M. 1993, \apj, 402, 76
 \reference{burs84} Burstein,~D., \& Heiles,~C. 1984, \apjs, 54, 33
 \reference{card89} Cardelli,~J.~A., Clayton,~G.C., \& Mathis,~J.S. 1989, 
\apj, 345, 245
 \reference{chio93} Chiosi,~C., Wood,~P.~R., \& Capitanio,~N. 1993, 
\apjs, 86, 541
 \reference{cowa94} Cowan,~J.~J., Romanishin,~W., \& Branch,~D. 1994, 
\apj, 436, L139
 \reference{deva91} de Vaucouleurs,~G., de Vaucouleurs,~A., Corwin,~H., 
Buta,~R., Paturel,~A., \& Fouqu\'{e},~P.  1991, in Third Reference Catalog
of Bright Galaxies, (Berlin: Springer) (RC3)
%\reference{dufo90} Dufour,~R. 1990, in Symp. on Evolution in Astrophysics, 
ESA SP-310
 \reference{ferr96} Ferrarese,~L. et al.\ 1996, \apj, 464, 568
 \reference{fish81} Fisher,~J.~R., \& Tully,~R.~B. 1981, \apjs, 47, 139
 \reference{free85} Freedman,~W.~L. 1985, \apj, 299, 74
 \reference{free88} Freedman,~W.~L. 1988, \apj, 326, 691
 \reference{free90a} Freedman,~W.~L. 1990, \apj, 355, L35
 \reference{free90b} Freedman,~W.~L., \& Madore,~B.~F. 1990, \apj, 365, 186
 \reference{free94a} Freedman,~W.~L., et al.\ 1994a, \apj, 427, 628
 \reference{free94b} Freedman,~W.~L., et al.\ 1994b, \nat, 371, 757
 \reference{garc93} Garcia,~A.~M. 1993, \aaps, 100, 47
 \reference{goul94} Gould,~A.\ 1994, \apj, 426, 542
 \reference{grah97} Graham,~J.~A., et al.\ 1997, \apj, 477, 535
%\reference{harr93} Harris,~H.~C., et al.\ 1993, \aj, 105, 1196
 \reference{hill98} Hill,~R.~J., et al.\ 1998, \apj, in press
 \reference{hoag83} Hoaglin,~D.~C, Mosteller,~F., \& Tukey,~J.~W. 1983, 
in Understanding Robust and Exploratory Data Analysis, (New York: Wiley)
 \reference{holt95a} Holtzman,~J.~A., et al.\ 1995a, \pasp, 107, 156
 \reference{holt95b} Holtzman,~J.~A., et al.\ 1995b, \pasp, 107, 1065
 \reference{huch97} Huchra,~J. 1997, CfA Redshift Catalog, private 
communication
 \reference{huch89} Huchtmeier,~W.~K., \& Richter,~O.~-G. 1989, in A General 
Catalog of \ion{H}{1} Observations of Galaxies, (New York: Springer)
 \reference{hugh89} Hughes,~S.~M.~G. 1989, \aj, 97, 1634
 \reference{hugh94} Hughes,~S.~M.~G. et al.\ 1994, \apj, 428, 143
 \reference{hugh96} Hughes,~S., Han,~M., \& Hoessel,~J. 1996, in Science with 
HST - II, eds. P.~Benvenuti, F.~D.~Macchetto, \& E.~J.~Schreier 
(ESA/NASA Publication), p46
 \reference{kels96} Kelson,~D.~D. et al.\ 1996, \apj, 463, 26
 \reference{kenn95} Kennicutt,~R.~C., Freedman,~W.~L., \& Mould,~J.~R. 1995, 
\aj, 110, 1476
 \reference{kenn98} Kennicutt,~R.~C., et al.\ 1998, \apj, submitted
%\reference{kraa86} Kraan-Korteweg,~R.~C. 1986, \aaps, 66, 255
 \reference{lafl65} Lafler,~J., \& Kinman,~T.~D. 1965, \apjs, 11, 216
 \reference{lee93}  Lee,~M.~G., Freedman,~W.~L., \& Madore,~B.~F. 1993, 
\apj, 417, 553
%\reference{mado85} Madore,~B.F. 1985, in IAU Coll. 82, in Cepheids, Theory and
%Observations, ed B.~F.~Madore, (Cambridge: Cambridge University Press), p166 
 \reference{mado91} Madore,~B.~F., \& Freedman,~W.~L. 1991, \pasp, 103, 933 (MF91)
 \reference{mado82} Madore,~B.~F. 1982, \apj, 253, 575
 \reference{marc94} Marcelin,~M., Petrosian,~A.~R., Amram,~P., \& Boulesteix,~J.
1994, \aap, 282, 363
 \reference{mart75} Martin,~W.~L., Warren,~P.~R., \& Feast,~M.~W. 1979, 
\mnras, 188, 139
% \reference{moul80} Mould,~J., Aaronson,~M. \& Huchra,~J. 1980, \apj, 238, 458
 \reference{moul96} Mould,~J., Sakai,~S., Hughes,~S., \& Han,~M. 1996, in 
Proceedings of the STScI May Symposium on the Extragalactic Distance Scale 
(Cambridge: Cambridge University Press), p158
 \reference{oey93} Oey,~M.~S., \& Kennicutt,~R.~C. 1993, \apj, 411, 137
 \reference{ohya96} Ohyama,~Y., \& Taniguchi,~Y. 1996, in ASP Conf Ser 103, 
The Physics of Liners in View of Recent Observations, eds. M.~Eracleous, 
A.~Koratkar, C.~Leitherer, and L.~Ho, p205
 \reference{osma82} Osman,~A.~M.~I., Ella,~M.~S., \& Issa,~I.~A. 1982, AN, 303, 329
 \reference{phel98} Phelps,~R., et al.\ 1998, \apj, submitted
%\reference{pier92} Pierce,~M.~J., \& Tully,~R.~B. 1992, \apj, 387, 47
 \reference{prad96} Prada,~F. 1996, private communication
 \reference{raws97} Rawson,~D.~M., et al.\ 1997, \apj, 488, 000
 \reference{rubi65} Rubin,~V.~C., Burbidge,~E.~M., Burbidge,~G.~R., 
Crampin,~D.~J., \& Prendergast,~K.~H.  1965, \apj, 141, 759
 \reference{saha90} Saha,~A., \& Hoessel,~J.~G. 1990, \aj, 99, 97
 \reference{saha94} Saha,~A., et al.\ 1994, \apj, 425, 14
 \reference{saha96} Saha,~A., et al.\ 1996, \apj, 466, 55
%\reference{sand93} Sandage,~A. 1993, \apj, 404, 419
 \reference{sand81} Sandage,~A., \& Tammann,~G. 1981, in A Revised Shapley-Ames
Catalog of Bright Galaxies, (Washington DC: Carnegie Institution of 
Washington)
 \reference{sass97} Sasselov,~D.~D., et al.\ 1997, \aap, in press
 \reference{sche93} Schechter,~P.~L., Mateo,~M., \& Saha,~A. 1993, 
\pasp, 105, 1342
 \reference{schm94} Schmidt,~B.~P., et al.\ 1994, \apj, 432, 42
 \reference{sers60} Sersic,~J.~L. 1960, Z.f.Astrophys., 50, 168
 \reference{silb96} Silbermann,~N.~A., et al.\ 1996, \apj, 470, 1
 \reference{stel78} Stellingwerf,~R.~F. 1978, \apj, 224, 953
 \reference{stet92} Stetson,~P.~B. 1992, in Stellar Photometry -- 
Current Techniques and Future Developments, IAU Coll.\ 136, eds.\ C.~J.~Butler
and I.~Elliot, p.~291
 \reference{stet94} Stetson,~P.~B. 1994, \pasp, 106, 250
 \reference{stif95} Stift,~M.~J.\ 1995, \aap, 301, 776
 \reference{tanv95} Tanvir,~N.~R., et al.\ 1995, \nat, 377, 27
 \reference{tanv96} Tanvir,~N.~R. 1996, Proceedings
of the STScI May Symposium on the Extragalactic Distance Scale, 
(Cambridge: Cambridge University Press), in press
 \reference{tonr97} Tonry,~J.~L., Blakeslee,~J.~P., Ajhar,~E.~A., \& 
Dressler,~A. 1997, \apj, 475, 399
 \reference{torm95} Tormen,~G., \& Burstein,~D. 1995, \apjs, 96, 123
 \reference{tull88} Tully,~R.~B. 1988, Nearby Galaxies Catalog, 
(Cambridge: Cambridge University Press)
 \reference{vand60} van den Bergh,~S. 1960, Public. David Dunlap Observatory, 2, \#6
%\reference{walk94} Walker,~A.~R. 1994, \pasp, 106, 828
 \reference{wegn93} Wegner,~G., Haynes,~M.~P., \& Giovanelli,~R. 1993, 
\aj, 105, 1251
 \reference{welc93} Welch,~D.~L., \& Stetson,~P.~B. 1993, \aj, 105, 1813
 \reference{west97} Westerlund, B. E. 1997, The Magellanic Clouds, Cambridge 
Astrophysics Series: 29, (Cambridge: Cambridge University Press)
%\reference{will96} Willick,~J.~A. et al.\ 1996, \apj, 457, 460
 \reference{zari90} Zaritsky,~D., et al.\ 1990, \aj, 99, 1108
 \reference{zari94} Zaritsky,~D., Kennicutt,~R.~C., \& Huchra,~J.~P. 1994, 
\apj, 420, 87

\end{references}
\end{document}